\documentclass[preprints,review,accept,moreauthors,pdftex]{Definitions/mdpi}

\usepackage{amsmath, amssymb, dsfont}

\firstpage{1} 
\pubvolume{xx}
\issuenum{1}
\articlenumber{5}
\pubyear{2019}
\copyrightyear{2019}
\history{}

\pdfoutput=1

\def\nue{\ensuremath{\nu_{e}\ }}
\def\nubare{\ensuremath{\overline{\nu}_{e}}\ }

\def\numu{\ensuremath{\nu_{\mu}\ }}
\def\nubarmu{\ensuremath{\overline{\nu}_{\mu}}}

\def\nutau{\ensuremath{\nu_{\tau}\ }}

\newcommand{\thetaot}{\ensuremath{\theta_{13}}\,}
\newcommand{\thetatt}{\ensuremath{\theta_{23}}\,}
\newcommand{\dmatm}{\ensuremath{\Delta m^2_{23}}}
\newcommand{\numunue}{\ensuremath{\nu_\mu \rightarrow \nu_e\,}}
\newcommand{\numunutau}{\ensuremath{\nu_\mu \rightarrow \nu_\tau\,}}
\newcommand{\dmot}{\ensuremath{\Delta m^2_{12}\,}}
\newcommand{\dmtt}{\ensuremath{\Delta m^2_{23}\,}}
\newcommand{\deltacp}{\ensuremath{\delta_{CP} \,}}
\newcommand{\numunumu}{\ensuremath{\numu \rightarrow \numu\,}}

\newcommand{\nova}{NO$\nu$A$\,$}

\newcommand{\be}{\begin{equation}}
\newcommand{\eeq}{\end{equation}}
\newcommand{\bea}{\begin{eqnarray}}
\newcommand{\eea}{\end{eqnarray}}
\newcommand{\nn}{\nonumber}

\Title{Three-flavour oscillations with accelerator neutrino beams}

\Author{Mauro Mezzetto $^{1}$\orcidA{} and Francesco Terranova $^{2,}$*\orcidB}

\AuthorNames{Mauro Mezzetto and Francesco Terranova}

\address{%
$^{1}$ \quad INFN sez. di Padova, Via F. Marzolo 8, I-35131, Padova, Italy; mauro.mezzetto@pd.infn.it\\
$^{2}$ \quad Dep. of Physics, Univ. of Milano-Bicocca and INFN sez. di Milano-Bicocca, Piazza della Scienza 3, I-20126, Milano, Italy; francesco.terranova@mib.infn.it}

\corres{Correspondence: francesco.terranova@mib.infn.it}

\abstract{The three-flavor neutrino oscillation paradigm is well established in particle physics thanks to the crucial contribution of accelerator neutrino beam experiments. In this paper we review the most important contributions of these experiments to the physics of massive neutrinos after the discovery of \thetaot and future perspectives in such a lively field of research. Special emphasis is given to the technical challenges of high power beams and the oscillation results of T2K, OPERA, ICARUS and \nova. We discuss in details the role of accelerator neutrino experiments in the precision era of neutrino physics in view of DUNE and Hyper-Kamiokande, the programme of systematic uncertainty reduction and the development of new beam facilities.}


\begin{document}

\section{Introduction}

High intensity neutrino beams at energies above 100 MeV are produced at proton accelerators since 1962 and played a key role 
in the foundation of the electroweak theory and the study of nuclear structure. Unlike other neutrino sources, they provide an unprecedented level of control of the flavor at source, flux and energy, which has been prominent in confirming the very existence of neutrino oscillations and measuring with high precision the oscillation parameters. Accelerator neutrino beams~\cite{kopp2007} are mostly sources of $\nu_\mu$ beams
because the vast majority of neutrinos produced at accelerators originate by the decay in flight of multi-GeV pions. As a consequence, they play a role in neutrino oscillation physics only if the oscillation length $L_{o}$ is within the earth diameter for the neutrino energy range $E$ accessible to accelerators (0.1-100 GeV). The existence of at least one oscillation length smaller than the earth diameter and involving $\nu_\mu$ in the GeV-energy range has been established by Super-Kamiokande in 1998~\cite{Fukuda:1998mi},~\cite{pdg}. In the modern formalism that describes neutrino oscillation (see Sec.~\ref{sec:oscillations}) it corresponds to saying that {\it both the mixing angle between the second and third neutrino mass eigenstates (\thetatt) and their squared mass difference (\dmatm $\equiv m^2_3 - m_2^3$) are different from zero}. Such statement opened up the modern field of neutrino oscillations at accelerators. Since $\dmatm \simeq 2.3 \times 10^{-3}$ eV$^2$ and the oscillation length is
$$
L_o (\mathrm{km}) \simeq 2.48 E(\mathrm{GeV})/\Delta m_{23}^2 (\mathrm{eV}^2)
$$
oscillations are visible for a GeV neutrino if the source-to-detector distance (``baseline'', $L$) is $L \simeq L_0 \simeq \mathcal{O} (1000 \ \mathrm{km})$. It is fortunate that oscillations are visible at propagation lengths smaller than the earth diameter because both the source and the detector can be artificial, i.e. designed by the experimentalist to maximise sensitivity and precision. On the other hand,
it is unfortunate that the baselines of interest are hundreds of km long (``long-baseline experiments''): since neutrinos cannot be directly focused, the divergence of the $\nu$ beam is  large and the transverse size of the flux at the detector is much larger than the size of the detector itself. As a consequence, the number of neutrinos crossing the detector decreases as $L^{-2}$ and such a drop must be compensated by brute force, increasing the number of neutrinos at source. All beams discussed in this paper are high power beams driven by proton accelerators operating at an average power $>100 $kW.

Even if the inception of the study of massive neutrinos at accelerators can be dated back to 1998, the golden age of accelerator neutrino physics has just started, triggered by the discovery of \thetaot. In 2012 reactor (Daya-Bay \cite{daya_bay_2012}, RENO \cite{reno_2012}) and accelerator (T2K \cite{t2k_2013}) experiments
demonstrated that {\it even the mixing angle between the first and third neutrino mass eigenstate is non-zero}. Indeed, \thetaot is quite large ($\thetaot \simeq 8^\circ$) so that the entire neutrino mixing matrix (PMNS - see Sec.~\ref{sec:oscillations}) has a very different structure compared with the Cabibbo-Kobayashi-Maskawa matrix. This result corroborates once more the fact that the Yukawa sector of the Standard Model is far from being understood both in the quark and lepton sector. From the experimental point of view, a large \thetaot opens up the possibility of observing simultaneously  \numunutau and \numunue oscillations in long baseline experiments and, in general, three family interference effects. This is the reason why neutrino physicists refer to the current and future generation of beam experiments as the experiments for the ``precision era'' of neutrino physics. These modern experiments must disentangle the missing parameters of the SM Lagrangian in the neutrino sector (the CP violating phase, the ordering of the mass eigenstates) from small perturbations in the $\nu_\mu$ oscillation probability at the GeV scale. Since the baselines of these experiments range from 295 km (T2K) to 1300 km (DUNE), beam intensity and statistics still play an important role. However, systematics will soon become the limiting factor for precision neutrino physics, especially in the DUNE and Hyper-Kamiokande era. 

In this paper, we will review the most important results appeared after the discovery of \thetaot in long baseline beams of increasing power. We will start in Sec.~\ref{sec:cngs} from the CNGS, a 500~kW power beam operated up to 2012 to study \numunutau oscillations. Subleading \numunue oscillations are currently investigated by two very high power beams at J-PARC and Fermilab serving T2K (Sec.~\ref{sec:t2k}) and \nova (Sec.~\ref{sec:nova}) respectively. These acceleration complexes will be upgraded to the status of ``Super-beams'' ($>1$ MW power) to carry on the ambitious physics programme of DUNE and Hyper-Kamiokande (Sec.~\ref{sec:dune}). The ancillary systematic reduction programme using current and novel acceleration techniques will be discussed in Sec.~\ref{sec:sys}.

\section{Three family neutrino oscillations}
\label{sec:oscillations}

The Standard Model (SM) of the electroweak interactions has been developed when no evidence of massive neutrinos was available. In its original formulation, therefore, the SM neutrinos were neutral massless fermions with left-handed (LH) chirality or, equivalently, LH helicity. The antineutrinos were their right-handed (RH) partners. The discovery of neutrino oscillations falsified the SM in its original formulation although the theory can be minimally extended to account for massive neutral leptons. In the minimally extended SM \cite{GonzalezGarcia:2007ib}, massive neutral leptons are treated as (massive) quarks: their flavour eigenstates are linear combination of mass eigenstates and the linear operator that mixes the flavour and mass eigenfunctions is a $3\times 3$ complex matrix. In the quark sector, this matrix is called the Cabibbo-Kobayashi-Maskawa (CKM) matrix. The corresponding matrix in the lepton sector is the Pontecorvo-Maki-Nakagawa-Sakata (PMNS) matrix $U_{\alpha, i}$. The $\alpha$ index runs over the flavor eigenstates ($\alpha = e,\mu,\tau$) and the $i$ index runs over the mass eigenstates $i=1,2,3$. The flavor fields $\nu_{\alpha L} (x) \equiv \nu_{\alpha} $  are thus:

\be
  \nu_{\alpha} =\, \sum_i U_{\alpha i} \nu_{i}  ,
\label{eq:flavour_mass_relation}
\eeq
Here, we dropped the dependence of the field on space-time and the subscript $L$. Note that in the (minimally extended) SM
only fields with LH chirality ($\nu_L$) appear in the charged currents (CC) that describe the coupling of fermions with the $W^{\pm}$ bosons. 
In the original SM, $\nu_L$ describe neutrinos with LH chirality that can be observed as particles with LH helicity. In the minimally extended SM, neutrinos are massive and therefore neutrinos with LH chirality may be observed with both spin antiparallel  (LH helicity) and parallel (RH helicity)  to the direction of motion. Direct mass measurements and oscillation data, however, suggest $m_i \ll m_e$ by at least five order of magnitudes. As a consequence, for any practical purpose the helicity of a LH chirality field can be considered LH, as well.
The neutrinos fields in the SM are Dirac fields, as for the case of quarks. The anti-neutrino fields describe particles that are different
from the neutrinos (``anti-neutrinos'') and only RH anti-neutrinos couple with the W bosons in CC currents. Unlike quarks, where particles are electrically charged and must be different from antiparticles of opposite electric charge, neutrinos are neutral particles and the anti-particle fields may be the same as for the particles. In this case, the neutrino fields are Majorana fields and not Dirac fields. This subtlety is immaterial for massless neutrinos because the RH neutrino field can be identified with the RH anti-neutrino field without changing the value of any observable. In the extended SM, there are observables that are sensitive to whether the neutrino are standard Dirac particles or "Majorana particles" (i.e. elementary fermions that are described by Majorana fields), although the correspondent measurements are extremely challenging due to the smallness of $m_i$. The minimally extended SM Lagrangian is 
built applying the quark formalism to neutrinos and, hence, in this model neutrinos are Dirac particles. In this case, as for the CKM, the PMNS matrix is unitary ($U^\dagger U = \mathds{1} $) and can be parameterized \cite{pdg} by three angles and one complex phase. The parameterization that has been adopted for the PMNS put the complex phase in the 1-3 sector, i.e. in the rotation matrix between the first and third mass eigenstate:

\bea
  U\! & =\! & \left( \begin{array}{ccc} 1 & 0 & 0 \\ 0 & c_{23} & s_{23} \\ 0 & - s_{23} & c_{23} \end{array} \right)
    \left( \begin{array}{ccc} c_{13} & 0 & s_{13} e^{-i \delta_{\rm CP}} \\ 0 & 1 & 0 \\
    - s_{13} e^{i \delta_{\rm CP}} & 0 & c_{13} \end{array} \right)
    \left( \begin{array}{ccc} c_{12} & s_{12} & 0 \\ - s_{12} & c_{12} & 0 \\ 0 & 0 & 1 \end{array} \right)   \nn \\
  & =\! & \left( \begin{array}{ccc} c_{12} c_{13} & s_{12} c_{13} & s_{13} e^{-i \delta_{\rm CP}} \\
    - s_{12} c_{23} - c_{12} s_{13} s_{23} e^{i \delta_{\rm CP}} &
    c_{12} c_{23} - s_{12} s_{13} s_{23} e^{i \delta_{\rm CP}} & c_{13} s_{23} \\
    s_{12} s_{23} - c_{12} s_{13} c_{23} e^{i \delta_{\rm CP}} &
    - c_{12} s_{23} - s_{12} s_{13} c_{23} e^{i \delta_{\rm CP}} & c_{13} c_{23} 
   \end{array} \right) \ .
\label{eq:PMNS}
\eea
In Eq.~(\ref{eq:PMNS}), the three rotation angle are labeled $\thetaot, \thetatt, \theta_{12}$ and $c_{ij} \equiv \cos \theta_{ij}$, $s_{ij} \equiv \sin \theta_{ij}$. In this parameterization, which is in use since more than 20 years, $\theta_{ij} \in \left[ 0, \frac{\pi}{2} \right]$ and $\delta_{\rm CP} \equiv \delta \in \left[ 0, 2 \pi \right]$. Clearly, physical observables are parameterization-independent and therefore 
the range of the angles and the choice of the complex phase in the 1-3 sector is done without loss of generalization.
All parameters of the minimally extended SM can be completely determined by neutrino oscillation experiments except for an overall
neutrino mass scale (see Table~\ref{tab:parameters}). The neutrino oscillation probability between two flavour eigenstates $P (\nu_\alpha \to \nu_\beta)\ $ is the probability to observe a flavour $\beta$ in a neutrino detector located at a distance $L$ from the source. The source produces neutrinos with flavor $\alpha$ and energy $E$. The oscillation probability is given by

\bea
  P (\nu_\alpha \to \nu_\beta)\ = &\! \delta_{\alpha \beta}\!\! &
    -\ 4 \sum_{i < j}\, \mbox{Re} \left[ U_{\alpha i} U^*_{\beta i} U^*_{\alpha j} U_{\beta j} \right]
    \sin^2 \left( \frac{\Delta m^2_{ji} L}{4 E} \right)  \nn \\
    && +\ 2 \sum_{i < j}\, \mbox{Im} \left[ U_{\alpha i} U^*_{\beta i} U^*_{\alpha j} U_{\beta j} \right]
    \sin \left( \frac{\Delta m^2_{ji} L}{2 E} \right) ,
\label{eq:oscillation_probability}
\eea
where $\Delta m^2_{ji} \equiv m^2_j - m^2_i$ and $L \simeq c t$ is the distance
travelled by the neutrino. For antineutrino oscillations, $U$ must be replaced by
$U^*$ in eq.~(\ref{eq:oscillation_probability}), which correspond  to changing the sign
of the third term. As a consequence, oscillation experiments can reconstruct all rotation angles and the CP phase.
They also can measure the squared mass differences among eigenstates although the absolute mass scale must be measured in an independent
manner. Note in particular, that CP violation in the leptonic sector can be established just measuring the difference between 
 $P (\nu_\alpha \to \nu_\beta)\ $ and  P $(\bar{\nu}_\alpha \to \bar{\nu}_\beta)\ $ if $\alpha \neq \beta$. Since the leading oscillation terms, i.e. the second term of eq.~(\ref{eq:oscillation_probability}), depend on a squared sign, determining the signs of $\Delta m_{ij}$
is also very challenging. In fact, the sign of \dmot has been determined at the beginning of the century by solar neutrino experiments.
The oscillation probabilities here are strongly perturbed by matter effects in the sun, which are sensitive to the sign of \dmot. 
As discussed in Sec.~\ref{sec:nova}, matter effects on the earth are employed by long-baseline experiments to determine the sign
of \dmtt, which is not fully established yet.   

The choice of Dirac fields for the description of neutrinos in the minimally extended SM is the simplest from the technical point of view but it is very  disputable in the light of the gauge principle \cite{Altarelli:2004za}. The most general Lagrangian compatible with the $SU(2) \otimes U(1)$ gauge symmetry and  containing all SM particles (including massive neutrinos) allows for ``Majorana mass terms'' as $- \frac{1}{2}\, m_i\, \nu^T_{i L} C \nu_{i L} + \mbox{h.c.}$, where $C$ is the charge
conjugation matrix. It is possible to show that these terms imply neutrinos to be Majorana particles and must be removed by hand from the minimally extended SM based on Dirac fields. A vast class of SM extensions is motivated by the existence of Majorana particles.  Several of these models are also able to explain the smallness of neutrino masses through the ``see-saw mechanism'' \cite{Minkowski:1977sc,GellMann:1980vs,Yanagida:1980xy,Mohapatra:1979ia,Schechter:1980gr}. Neutrino oscillations, however,
cannot address these kind of issues. If neutrinos are Majorana particles, the PMNS matrix is only approximately unitary and can be parameterized with three
rotation angles and three phases as

\be
U = U_{\rm Dirac} P_{\rm Majorana}\
\eeq
where
\be
  P_{\rm Majorana}\, =\, \left( \begin{array}{ccc} e^{i \alpha_1} & 0 & 0 \\ 0 & e^{i \alpha_2} & 0 \\ 0 & 0 & 1 \end{array} \right)
\eeq
and $U_{Dirac}$ is the PMNS in the minimally extended SM (eq.~\ref{eq:PMNS}). On the other hand, the oscillation formula of eq.~\ref{eq:oscillation_probability} does not depend on these additional phases. The oscillation measurements are therefore robust against the Dirac vs Majorana nature of neutrinos but, on the other hand, do not provide information to determine such nature.  
        
\begin{table}[H]
\caption{Current value of the mass and mixing parameters as extracted from neutrino oscillation data under the hypothesis of normal hierarchy (current best fit)  \cite{Esteban:2018azc}.}
\centering
\begin{tabular}{ccc}
\toprule
\textbf{parameter}	& \textbf{best estimate} & \textbf{units} \\
\midrule
$\theta_{12}$   & $ 33.82^{+0.78}_{-0.76}$ & degree \\ 
\vspace{0.2cm} $\theta_{23}$   & $ 48.3^{+1.1}_{-1.9}$ & degree \\
\vspace{0.2cm} $\theta_{13}$   & $ 8.61 \pm 0.13$ & degree \\
\vspace{0.2cm} $\delta$   & $ 222 ^{+38}_{-28}$ & degree \\
\vspace{0.2cm} \dmot   & $ 7.39 ^{+0.21}_{-0.20} \times 10^{-5}$ & eV$^2$ \\
\vspace{0.2cm} \dmtt   & $ 2.52 \pm 0.03 \times 10^{-3}$ & eV$^2$ \\
\bottomrule
\end{tabular}
\label{tab:parameters}
\end{table}

\section{The CNGS neutrino beam}
\label{sec:cngs}

In 1998, disappearance measurements of atmospheric muon neutrinos - i.e. measurements of \numunumu with neutrinos produced by interactions of cosmic rays with the earth atmosphere - provided evidence for non-zero value of both \dmtt and \thetatt. At that time, \thetaot was not known and only upper limits from the Chooz reactor experiment~\cite{Apollonio:2002gd} were available ($\thetaot < 12^\circ$). In those years, there was no evidence of oscillations in appearance mode, i.e. no measurements testifying non-zero values of $P (\nu_\alpha \to \nu_\beta)\ $ for any $\alpha \neq \beta$. Such evidence was still lacking in 2002, when solar and reactor experiments employing \nue and \nubare whose energy was well below the kinematic threshold for the production of muons, demonstrated that $\theta_{12} \simeq 35^\circ$ and
$\dmot \simeq 7 \times 10^{-5}$ eV$^2$. In order to observe in a neutrino detector the appearance of new flavors ($\alpha \neq \beta$), the energy of the neutrino must be well above the kinematic threshold for the production of the lepton of flavor $\beta$. For solar and reactor neutrinos this is not possible since the energies of their \nue and \nubare are much smaller than $m_\mu$. Atmospheric neutrinos, on the other hand, may be the right natural source since their energy spectrum extend well above the $\mu$ and even $\tau$ kinematic threshold. At that time, there was no evidence of \numunue oscillations at the atmospheric energy scale and, therefore, the disappearance of \numu was ascribed to \numunutau oscillations. The CERN-to-Gran Sasso (CNGS) \cite{cngs,Gschwendtner:2013mma} neutrino beam was designed to provide evidence of \numunutau oscillations in appearance mode, i.e. observing explicitly the appearance of \nutau by detecting $\tau$ leptons produced by CC events in the detector. The CNGS beam exploited the Gran Sasso laboratories (LNGS), an existing underground laboratory located 730~km from CERN. This constraint fixed the baseline of the CNGS experiments: OPERA and ICARUS.The CNGS has been the only long-baseline beam whose energy ($E\simeq 17$~GeV) was well above the kinematic threshold for the production of the tau lepton (3.5 GeV for CC scattering in nuclei). The energy was tuned maximizing the number of observable \nutau CC events:

\be
\int dE \Phi(E) \epsilon (E) \sigma(E) P(\numunutau) \simeq \sin^2 2\thetatt \int dE \Phi(E) \epsilon (E) \sigma(E) \sin^2 \left( 1.27 \ \frac{ \dmtt \cdot 730 \ \mathrm{km} }{E (\mathrm{GeV})} \right)
\label{eq:cngs_events}
\eeq     
where $\Phi(E)$ is the CNGS $\numu$ flux, $\epsilon(E)$ is the detector efficiency and $\sigma(E)$ is the \nutau CC cross section. 
In this formula, we expressed the oscillation phase $\dmtt L/4E$ in practical units: km for $L$ and GeV for $E$.
In particular, a high energy beam  is beneficial to produce tau leptons with a large cross section ($\sigma(E) \sim E$ for $E$ well above 3.5 GeV) and boosted in the laboratory frame. Such boost is needed to observe the decay in flight of the taus in the OPERA nuclear emulsion sheets (see below) and increases substantially $\epsilon(E)$.
On the other hand, an energy much larger than $1.27 \cdot \dmtt \cdot 730 \ \mathrm{km}$  reduces the oscillation probability as $\sim E^{-2}$ due to the fixed baseline of LNGS.
The actual CNGS beam energy has thus been optimized accounting for these requirements and 17 GeV is the value that maximises Eq. \ref{eq:cngs_events}.
As a matter of fact,
the CNGS observes oscillations off the peak of the oscillation maximum since 

\be
  \sin^2 \left( 1.27 \ \frac{ \dmtt \cdot 730 \ \mathrm{km} }{17 \ \mathrm{GeV}} \right) \simeq 0.016
\eeq
In the framework of the minimally extended SM, the CNGS thus provides a direct test of the oscillation phenomenon in appearance mode (\numunutau) and a test of the oscillation pattern in a parameter phase far from the oscillation maximum. 

\begin{figure}
\centering
\includegraphics[width=0.8\textwidth]{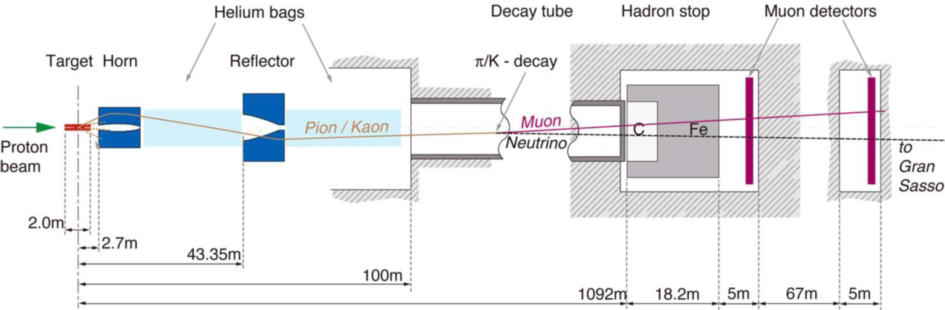}
\caption{Layout of the CNGS neutrino beam.}
\label{fig:cngs_layout}
\end{figure}

The CNGS neutrino beam (see Fig.~\ref{fig:cngs_layout}) was produced by 400 GeV/c protons extracted from the SPS
accelerator at CERN and transported along a 840 m long beam-line to a carbon target producing kaons and
pions \cite{opera_detector}. The only potential source of $\nutau$ were the decay of $D_s$ particles in the proximity of the target, which contributed in a negligible manner (<10$^{-4}$) to the neutrino flux. The positively charged $\pi$/K were energy-selected and guided with two focusing lenses (``horn'' and ``reflector''), in the direction of the Gran Sasso Laboratories in Italy. These particles decay into $\mu$  and \numu in a
1 km long evacuated decay pipe. All the hadrons, i.e. protons that have not interacted in the target, pions
and kaons that have not decayed in flight, are absorbed in a hadron stopper. Only neutrinos and
muons cross the 18 m long block of graphite and iron. The muons, which are ultimately
absorbed downstream in around 500 m of rock, were monitored by muon detector stations.

Unlike all other long baseline facilities, the CNGS was not equipped with a near detector due to the cost and complexity of building
a large underground experimental area located after the CNGS decay pipe. In addition, since the CNGS was designed to study the appearance of tau neutrinos
in a beam with a negligible contamination of $\nu_\tau$, the need for a near detector was less compelling than other long-baseline beams. 
The CNGS beam ran from 2008 to 2012, providing  to OPERA an integrated exposure of
$17.97 \times 10^{19}$ protons on target (pot). OPERA exploited this large flux using real-time (``electronic'') detector interleaved
with emulsion cloud chambers (ECC). 

The OPERA \cite{opera_detector} target had a total mass of about 1.25 kt composed of two
identical sections. Each section consisted of 31 walls of Emulsion
Cloud Chamber (ECC) bricks, interleaved by planes of horizontal and
vertical scintillator bars (Target Tracker) used to select ECC bricks
in which a neutrino interaction had occurred. Each ECC brick consisted
of 57 emulsion films interleaved with lead plates (1 mm thickness),
with a ($12.7 \times 10.2$) cm$^2$ cross section and a total thickness
corresponding to about 10 radiation lengths. Magnetic muon
spectrometers were located downstream of each target section, and were
instrumented by resistive plate chambers and drift tubes.

Over the years, a total of 19505 neutrino interaction events in the target were recorded by
the electronic detectors, of which 5603 were fully reconstructed in the OPERA emulsion films \cite{Agafonova:2019npf}.  

The construction and data analysis of OPERA was a tremendous challenge
due to the need of identifying tau leptons on an-event-by-event
basis. Even with current technologies, the only mean to achieve this
goal by detecting the decay kink with very high purity is by using nuclear emulsions. This technique has been exploited in the 90s by the CHORUS experiment to search for short baseline \numunutau oscillations \cite{Eskut:2007rn} and extended to a much larger scale by OPERA.  
Nuclear emulsions consist of AgBr crystals scattered in a gelatine
binder. After the passage of a charged particle, the crystal produces
electron-hole pairs proportional to the deposited energy. During a
chemical-physical process known as development, the reducer in the
developer gives electrons to the crystal through the latent image
center and creates silver metal filaments scavenging silver atoms from
the crystal. This process increases the number of metal silver atoms
by many orders of magnitude (10$^8$ - 10$^{10}$). The grains of silver
atoms reach a diameter about 0.6 mm and, therefore, become visible
with an optical microscope.
 
The development procedure is irreversible but it must be performed in
a finite time scale (a few years) to prevent aging of the
emulsions. As a consequence, the electronic detectors must locate the
ECC brick in real time and a semi-automatic system must remove the
brick while OPERA is still taking data and bring this component to the
development facility. The scanning of the emulsions can be done at any time
since the particle tracks are permanently fixed on the bulk of the emulsions but a fast scanning
time is a major asset to provide feedback on the detector performance and optimize the analysis chain.

From the analysis point of view, OPERA is a rare event search experiment since just an handful of
fully reconstructed tau events are expected from eq.~\ref{eq:cngs_events}. The tau events in OPERA are reconstructed as
events with a primary track that shows a kink with respect to the initial direction. The kink testifies the decay in flight
of the tau lepton in the lead of the ECC.
In particular, OPERA has been seeking
for tau leptons in most of its high branching ratio decay modes: $\tau \rightarrow \mu$ (i.e. $\tau \rightarrow \mu \numu \nutau$ with
a visible muon from the primary vertex), $\tau \rightarrow e$, $\tau \rightarrow h$ (``one prong'', i.e. hadronic decay with only one visible charged hadron) and ``three prongs'' (i.e. hadronic decay with three visible charged hadrons).  
Rare event searches are plagued by potential biases due to the tuning of the selection procedure in the course of data taking. In order to minimize bias, the OPERA Collaboration followed two approaches. The most conservative one was implemented during data taking and the initial stage of data analysis, bringing to a $5 \sigma$ discovery of tau appearance in 2015. Here, tau leptons were selected using topological and kinematic cuts defined a priori in the design phase of the experiment. The efficiency of these cuts were evaluated by a full simulation
of OPERA performed with GENIE~\cite{genie} (neutrino interactions), NEGN~\cite{negn} (kinematic properties of final states) and GEANT4~\cite{geant4} (detector response). The simulation was validated with a control sample based on $\numu$ CC events producing charmed particles in the final state. 
The 2015 analysis \cite{Agafonova:2015jxn} is based on 5 \nutau CC candidates whose background is summarized in table~\ref{tab:opera_background}.

\begin{table}[H]
\caption{Summary of expected events and background contributions in the 2015 OPERA analysis}
\centering
\begin{tabular}{cccc}
\toprule
\textbf{Decay mode}	& \textbf{Exp. Bkg}	& \textbf{Exp. Signal} & \textbf{Observed}\\
\midrule
$\tau \rightarrow h$		& 0.04 $\pm$ 0.01 & 0.52 $\pm$ 0.10 & 3 \\
$\tau \rightarrow 3h$		& 0.17 $\pm$ 0.03 & 0.75 $\pm$ 0.14 & 1 \\
$\tau \rightarrow \mu$		& 0.004 $\pm$ 0.001 & 0.61 $\pm$ 0.12 & 1 \\
$\tau \rightarrow e$		& 0.03 $\pm$ 0.01 & 0.78 $\pm$ 0.16 & 0 \\
\midrule
total		& 0.25 $\pm$ 0.05 & 2.64 $\pm$ 0.53 & 5 \\
\bottomrule
\end{tabular}
\label{tab:opera_background}
\end{table}

\begin{figure}[H]
\centering
\includegraphics[width=0.8\textwidth]{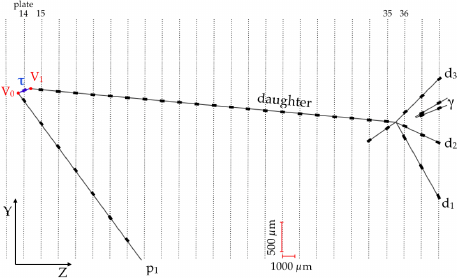}
\caption{The 5th $\tau$ candidate ($\tau \rightarrow h$) observed in one of the OPERA ECC in 2015.}
\label{fig:opera_5th}
\end{figure}

The events observed in OPERA for the standard 2015 analysis are consistent with the "three flavor neutrino oscillation paradigm", i.e. the minimally extended SM accounting for neutrino oscillations.
In particular, given the superior background rejection capability of the ECC based analysis, $\nutau$ appearance was established with a statistical significance of 5.1 $\sigma$ with just 5 candidates (see Fig.~\ref{fig:opera_5th}). For that particular analysis, the probability of detecting 5 events or more is 6.4\%. 

In 2018, OPERA developed a more sophisticated analysis reaping on the
improvement in the knowledge of the detector response and background
subtraction techniques accumulated during data taking. The final OPERA
analysis is based on looser kinematic cuts applied to four variables: the
distance  between the decay vertex and
the down-stream face of the lead plate containing the primary vertex ($z_{dec}$),
the momentum ($p$) and transverse momentum ($p_T$) of the secondary particle and the
3D angle $\theta_{kink}$ between the parent particle and its daughters.
The cuts are independent of the tau decay mode ($z_{dec}<2.6$~mm, $\theta_{kink}>0.02$~rad, $p>1$~GeV) except for the
transverse momentum ($p_T> 0.15$~GeV for $\tau \rightarrow h$ and  $p_T> 0.1$~GeV for $\tau \rightarrow e$ or $\mu$).
In addition $p<15$ GeV is requested for the tau decay into muon to suppress the high energy tail of $\numu$ CC events of the
CNGS. After these cuts, a multivariate statistical approach is employed to separate signal from background and to produce a single-variable discriminant for
the oscillation hypothesis test and parameters estimation.
In this analysis 10 observed events were selected. The expectation in the minimally extended SM was $6.8 \pm 0.75$ events plus a background of
$2.0 \pm 0.4$ events. This analysis, which represents the final result of OPERA on search for \numunutau appearance, establishes oscillations at 6.1 $\sigma$ confidence level \cite{Agafonova:2018auq}. 

Beyond tau appearance, OPERA performed analyses in $\nu_\mu$ disappearance mode and in $\numunue$ appearance mode. Even if these searches are not
competitive with dedicated experiments equipped with a near detector and electron-sensitive far detectors with much larger mass than OPERA (see Sec.\ref{sec:t2k} and \ref{sec:nova}), the three neutrino analysis of OPERA  \cite{Agafonova:2019npf} provides a consistency test of the minimally extended SM in an energy range between the atmospheric oscillation scale (few GeV neutrinos) and the energy scale of neutrino telescopes like IceCube/DeepCore. The OPERA and ICARUS (see below) experiments, in addition, provided significant constraints to the existence of light sterile neutrinos and, in general, to extension of the three family oscillation mechanism.

\begin{figure}
\centering
\includegraphics[width=0.7\textwidth]{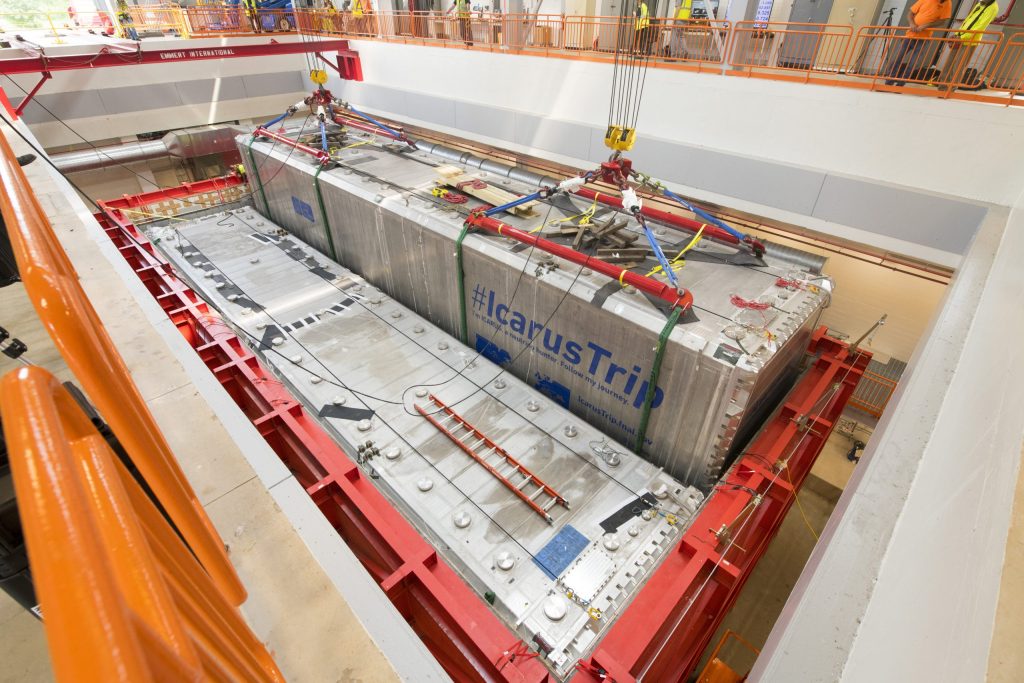}
\caption{The ICARUS detector, operated underground at LNGS during the run of CNGS, was moved  to Fermilab in 2018 and installed at a shallow depth to search for sterile neutrinos in the framework of the Fermilab Short Baseline Programme (SBN) \cite{Antonello:2015lea}.}
\label{fig:icarus}
\end{figure}

The CNGS served for several years also the ICARUS experiment \cite{Amerio:2004ze} that took
data at LNGS from 2010 to 2012.  ICARUS was a 600 ton liquid Argon TPC
(see Fig.~\ref{fig:icarus}) composed of a large cryostat divided into
two identical, adjacent half-modules, each one with internal
dimensions of 3.6~m (width) $\times$ 3.9~m (height) $\times$ 19.6~m
(length). The inner detector structure of each submodule (300 ton)
consisted of two TPCs separated by a common cathode.  Each TPC is made
of three parallel wire planes, 3 mm apart and the relative angles
between the two wire planes is 60$^\circ$.  The distance between the
wires is 3 mm (wire pitch). These wire planes constitute the anode of
the TPC. The TPC volume is filled with high purity liquid Argon and
the electrons produced by the charged particles drift toward the anode
wires. A uniform electric field perpendicular to the wires drives the
electrons in the LAr volume of each half-module by means of a HV
system. The cathode plane, parallel to the wire planes, is placed in
the center of the LAr volume of each half-module at a distance of
about 1.5 m from the wires of each side. This distance defines the
maximum drift path. Between the cathode and the anode field-shaping
electrodes are installed to guarantee the uniformity of the field
along the drift direction. At the nominal voltage of 75 kV,
corresponding to an electric field of 500 V/cm, the maximum drift time
in LAr is about 1 ms. As a consequence, the purity of the liquid Argon
must be such that the electron lifetime is comparable with the drift
time, i.e. that electrons can drift in the active volume before
recombining with electronegative impurities. Such a long drift length
requires Argon purity $<0.01$ part-per-billion (ppb). The space resolution that can be achieved in ICARUS is much worse than the ECC
(1 mm versus a few microns) but this detector technology can be readout
as a standard TPC and the observation and selection of CC events can
be performed in real time. In addition, this technology is scalable to
masses that are unpractical for the ECC ($\gg 1$~kton).

In fact, the mass of the TPC operated at LNGS - the largest LArTPC
ever built - was still smaller than what needed for the
observation of \nutau CC events on kinematical basis. Still, ICARUS
achieved important results in the study of \numunue oscillations
beyond the minimally extended SM (limits to sterile neutrino
oscillations) \cite{Antonello:2013gut} and in the clarification of the neutrino velocity
anomaly reported in 2011 by OPERA \cite{Antonello:2012hg}.  Even more, the successful
operation of ICARUS in a deep underground laboratory at the 600 t
scale, opened up the possibility of building multi-kton detectors
based on the LArTPC technology, paving the way to the design and
construction of ProtoDUNE-SP and DUNE (see Sec.~\ref{sec:dune}). At the end of the ICARUS data taking at LNGS, the Argon
recirculation and purification systems reached a free electron lifetime exceeding 15 ms,
corresponding to about 20 parts per trillion of O$_2$-equivalent
contamination \cite{Antonello:2014eha}.

\section{T2K and  its upgrades}
\label{sec:t2k}

\subsection{T2K}
The T2K experiment is the successor of the former K2K experiment, \cite{Oyama:1998bd}, the first long-baseline neutrino experiment 
that confirmed the 
Super-Kamiokande discovery of neutrino oscillations \cite{Fukuda:1998mi}
by detecting a reduction of the $\nu_\mu$ flux together with a distortion of the energy spectrum in an accelerator neutrino beam generated at KEK \cite{Ahn:2002up}.

With respect to K2K, T2K had a number of substantial improvements, the most important being a) a 
new neutrino beam created by the Japan Proton Accelerator Research Complex (J-Parc), designed to be about 50 times 
more intense than the K2K neutrino beam; b) a beam line configuration where the far detector (again 
SK) was at an off-axis angle of $2.5^\circ$ and a baseline of 295 km; c) a new close detector system, allowing a much more 
effective measurement of neutrino beam components, backgrounds and neutrino cross sections.

The T2K neutrino beam is generated at J-PARC where
a 30 GeV proton beam impinges onto a graphite target \cite{Abe:2011ks}.
The primary proton beam recently achieved an average power of 492 kW.
The secondary hadrons, mostly pions, are charge-selected and focused by a system of three magnetic horns \cite{Sekiguchi:2015ghw} and decay in a 96 m long decay volume. 
 The axis of the beam is displaced by $2.5^\circ$ from the SK detector.
 
Thanks to the kinematics of the pion two-body decay, the resulting neutrino beam is narrower than an on-axis beam \cite{E889}, better optimized to the expected neutrino oscillations (as already measured by SK and K2K), and with a smaller contamination of intrinsic \nue (mostly generated by kaon decays, whose kinematics is different from pion decays, producing less neutrinos at the off-axis angle of $2.5^\circ$).

Two detectors are located 280 m downstream of the beam target, to measure neutrino interactions before they start oscillating. The INGRID detector \cite{Abe:2011xv} is located on-axis and its main purpose is to monitor the direction and the flux stability of the neutrino beam.

The ND280 detector system 
is located as the same off-axis angle of SK and its main purpose is to measure all the 
neutrino beam components. It consists of of three time projection chambers (TPCs) \cite{Abgrall:2010hi} 
interleaved with two fine-grained tracking detectors \cite{Amaudruz:2012agx}, a $\pi ^\circ$-optimized detector (P0D) \cite{Assylbekov:2011sh} and a  
electromagnetic calorimeter \cite{Allan:2013ofa}. 
ND280 is mounted inside the magnet of the former UA1 and Nomad experiments, donated by CERN to the T2K Collaboration, 
it  allows to independently measure both neutrino and antineutrino interaction 
rates. A side muon range detector instruments the magnet yoke \cite{Aoki:2012mf}.

An important set of external data for T2K comes from the NA61 experiment 
that measured hadroproduction  by 30 GeV protons on  a graphite target
with a thickness of 4\% of a nuclear interaction length \cite{Abgrall:2015hmv}. These measurements drastically reduce 
systematic errors on the evaluation of the neutrino beam flux, resulting into a better determination of neutrino interaction 
rates by ND280 \cite{Abe:2012av}. 

The neutrino beam is peaked around 0.6 GeV, at the energy where the oscillation probability is maximal.
At those energies the dominant process are Charged Current Quasi-Elastic scatterings (CCQE),
 where a charged lepton of the same flavor of the incoming neutrino is produced.
In this class of interactions the measurement of the 
momentum and direction of the outgoing lepton is sufficient to reconstruct the energy of the incoming 
neutrino.
It has to be noted that the 295 km baseline is too short to allow good sensitivity to the measurement of neutrino mass ordering (see Sec. \ref{sec:nova}) and that the $\nu_\mu$ neutrino energy is well below the $\tau$ production threshold.

The far detector is the Super-Kamiokande (SK) detector \cite{Fukuda:2002uc}, a deep underground 50 kton water Cherenkov 
detector,  39 m in diameter and 42 m tall, equipped with 11,129 inward facing 20-inch  photomultiplier tubes (PMTs) and 1,885 outward-facing 8-inch PMTs mainly used as a veto.
The Cherenkov light emitted in water by charged particles above a momentum threshold allows to measure with great precision the momentum, the direction and the  vertex of the leptons produced by neutrino interactions. The different topologies of the muon and electron tracks produce different 
pattern of hit phototubes, in this way electron and muons are identified  with great 
precision by SK. The quasi-elastic topology, where a single lepton is produced, is particularly favorable 
for the characteristics of the SK detector.

A first important result by T2K was published in 2011 \cite{Abe:2011sj}  by observing an indication of 
\numunue appearance in data accumulated with $1.43 \cdot 10^{20}$ proton on target collected from 
January 2010 to March 2011. Six events passed all the selection criteria in SK, with an expected 
background of $1.5\pm 0.3$ events, corresponding to a significance of about $2.5 \sigma$. It represented the first 
indication of a non-zero value of \thetaot: $0.03 < \sin^2{2\thetaot} < 0.28$ at 90\% CL for \deltacp=0 and normal neutrino mass hierarchy, a fundamental milestone for any subsequent development in neutrino oscillation physics.

T2K had a long stop of more than one year soon after this result due to the catastrophic earthquake-tsunami that occurred in Japan in March 2011. 

\begin{figure}
\centering
\includegraphics[width=0.7\textwidth]{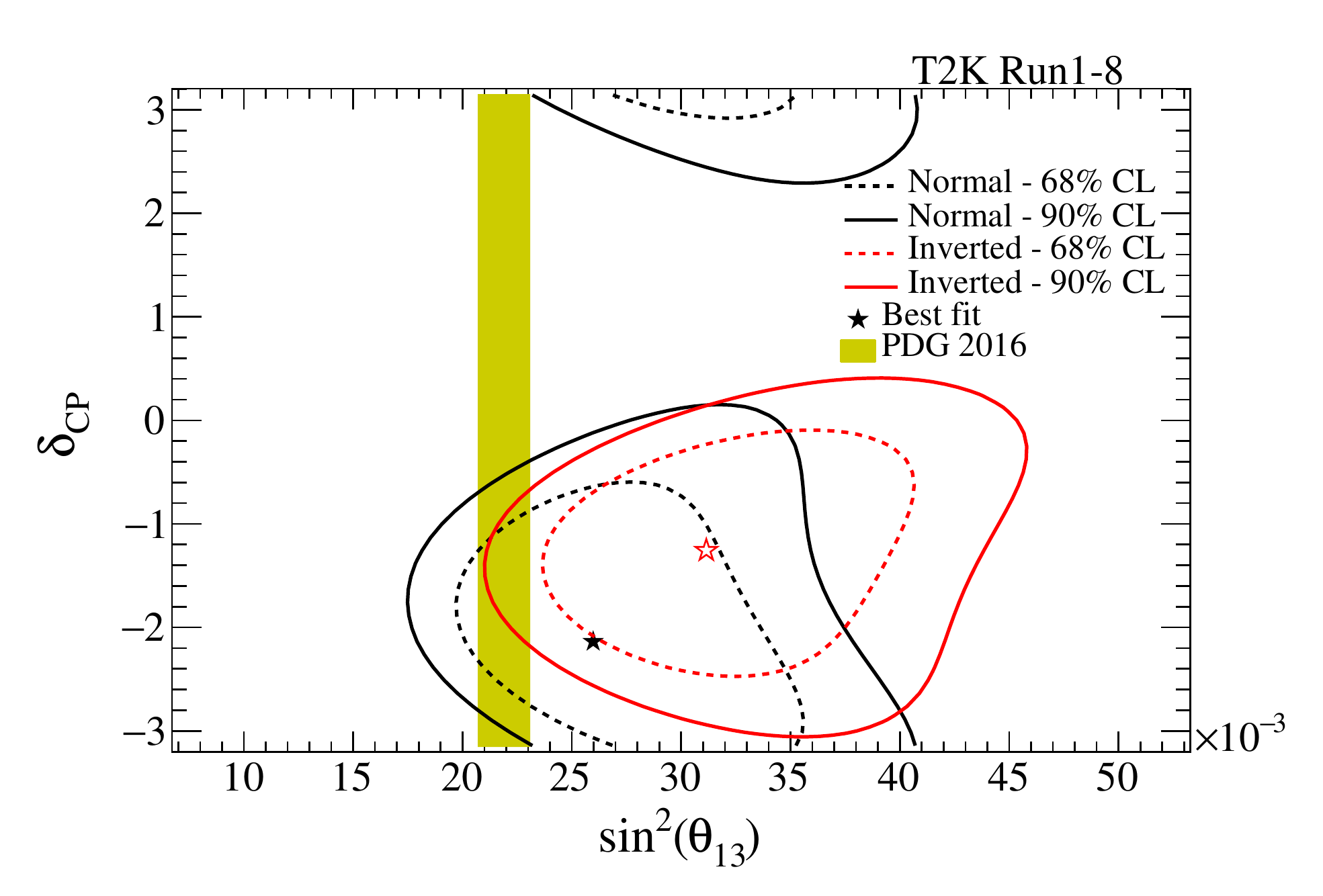}
\caption{Confidence regions (68\% and 90\%) in the $\sin^2{\theta_{13}}-\delta_{CP}$ plane as measured by T2K without any prior information on $\theta_{13}$ compared with the 68\% confidence regions from reactor experiments, shown by a yellow vertical band.}
\label{fig:T2K-dCP}
\end{figure}

In this period three reactor experiments published 
evidence of non-zero values of \thetaot by measuring the flux of reactor antineutrinos with a clear 
indication of \nubare disappearance. Eventually the \thetaot determination by reactor experiments 
achieved a much better precision than T2K and at today the global best-fit to \thetaot is \cite{pdg} 
$\sin^2{\thetaot} = 0.0215$ with a $3\sigma$ allowed range of $0.0190 - 0.0240$.
The precise, unambiguous, determination of \thetaot by reactors greatly improved the sensitivity of 
long baseline experiments to \deltacp and neutrino mass ordering.
This is nicely illustrated in Fig.\ref{fig:T2K-dCP} where the confidence regions measured by T2K without any input from reactor experiments are compared with the confidence region from reactors \cite{Abe:2018wpn}, indicating how better $\delta_{CP}$ is constrained after the inclusion of reactor data.

The most recent T2K published results are based on SK data collected between 2009 and 2017 in (anti)neutrino mode 
with a neutrino beam exposure of $14.7\times 10^{20}$ ($7.6\times 10^{20}$) protons on target 
\cite{Abe:2018wpn}.

Neutrino fluxes, interaction cross-sections and detector responses are simulated and
a fit to the near detector data constrains them.
The results of this fit are propagated to the far detector as a Gaussian covariance on
the systematic parameters. The fit of the far detector data then constrains the oscillation
parameters.

Primary proton interactions on the T2K target system are simulated with
FLUKA2011 \cite{Ferrari:2005zk} and GEANT3 \cite{Brun:1994aa}
together with the decay of hadrons and muons that produce neutrinos.
The modeling is constrained by thin target hadron production
data, including recent charged pion and kaon measurements from the NA61/SHINE experiment
 at CERN \cite{Abgrall:2011ae,Abgrall:2011ts,Abgrall:2015hmv}.
The systematic errors of beam prediction at this stage are of 8\% and 7.3\% for the neutrino
 and antineutrino beams respectively.
It's expected a significant reduction of these errors with the inclusion of
recent hadroproduction measurement of the T2K replica target by NA61/SHINE \cite{Berns:2018tap}.
 It is extremely important to have flux systematics as low as possible
 at this stage for any physics analysis at the close detector and to attenuate the strong
 correlation between neutrino fluxes and neutrino cross sections at the close detector.

The neutrino interaction model is based on NEUT \cite{Hayato:2009zz}. 
Simulated data is also generated with  alternative models 
and fitted with NEUT in order to evaluate the systematic effects of
wrong modelling to the oscillation results.

Events selected as charged
current (CC) interactions in ND280 are used to
constrain the flux and neutrino interaction uncertainties.

A total of 14 event samples are defined: 6 neutrino samples depending of the number of hadrons (0, 1 or more) and weather the interaction happen in FGD1 or FGD2, this latter being 40\% enriched in water;
8 samples are defined in the antineutrino mode, depending of the number of tracks (0 or 1), the charge of the outgoing muon and again on the position of the interaction vertex.

It is very important to have the possibility of defining such a large number of event samples, they are sensitive to different combinations of beam fluxes and neutrino interactions, helping to cover all the analysis parameters and  to break parameter degeneracies and ambiguities.

A global likelihood is computed for the fit of all the event samples. The global fits greatly reduces systematic errors and increases by 10\%-15\% the interaction rate predictions
at the far detector.

SK data is subdivided in five independent samples: events with a single muon-like or a single electron-like rings in neutrino and antineutrino beam modes plus a sample with an electron-like ring together with a pion track candidate in neutrino mode.
Events are reconstructed with a new reconstruction algorithm, firstly introduced in \cite{Abe:2015awa} to suppress neutral current $\pi^\circ$ backgrounds, where a maximul likelihood algorithm exploiting the timing and charge information of all the photosensors simultaneously. This new alghorithm improves efficiency and purity of the selected samples and allows a fiducial volume expansion of
25\% (14\%) for \nue (\numu) events.

The observed number of CCQE events with an electron candidate at SK are 74 in the neutrino run and 7 in the anti-neutrino run, to be compared
 with the expectations for $\deltacp=0 \, (-\pi/2)$, normal hierarchy, of 61.4 (73.5) and 9.0 (7.9) respectively.
 The systematic errors of the two samples are 8.8\% and 7.1\% respectively.
 The selected \numu candidates in the neutrino and antineutrino modes are 240 and 68 respectively, with 5.1\% and 4.3\% systematic errors.
 Finally 15 events are selected in the single-electron single-pion sample, with a prediction of 6.9 events and a systematic error of 18.4\%.
 It should be noted that systematic errors already at this stage (29\% of the design statistics) are sizable and are approaching the values of statistical errors.

Oscillation fits to the 5 samples of data
show a preference for the normal mass ordering with a posterior probability of 87\%. 
The oscillation angle $\theta_{23} $ is measured as $\sin^{2}(\theta_{23})= 0.526^{+0.032}_{-0.036}$ for normal ordering, central value for inverted ordering is $\sin^{2}(\theta_{23})= 0.530$. 
Assuming the normal (inverted) mass ordering the best fit to the atmospheric mass splitting  is 
$\Delta m^{2}_{32}= (2.463\pm 0.07)\times 10^{-3}$ $(\Delta m^{2}_{13}=(2.432\pm 0.07)\times 10^{-3})$ 
$\text{eV}^{2}/c^{4}$. 
For $\delta_{CP}$ the best-fit value  assuming the normal (inverted) 
mass ordering is $-1.87 (-1.43)$, the $2\sigma$ confidence regions are $(-2.99, -0.59)$ for normal ordering and $(-1.81, -1.01)$ for inverted ordering.
Both intervals rule out the CP conserving points,  $\delta_{CP}=0$ and $\delta_{CP}=\pi$.

A recent preprint of the T2K collaboration, which includes 2018 data and improves by a factor 2.2 the p.o.t. collected in antineutrino mode,
 reports for the first time a   closed 99.73\% ($3\sigma$) interval on the $CP$-violating phase $\delta_{CP}$ \cite{Abe:2019vii} . 

 Other notable results of the T2K Collaboration are several measurements of neutrino cross sections \cite{Avanzini:2018oie}, searches for light sterile neutrinos \cite{Abe:2019fyx} and searches for heavy neutrinos \cite{Abe:2019kgx}.

\subsubsection{T2K II}
\label{sec:T2K2}
The exciting results achieved by the T2K experiment so far, convinced the collaboration to propose an upgrade of the experimental setup, T2K II \cite{Abe:2016tez}, in order to reach a 3 $\sigma$ sensitivity for a significant fraction of the possible $\delta_{CP}$ values by the year 2026, when the next generation Hyper-Kamiokande experiment is expected to start. 

The proposed upgrades, all re-usable by the Hyper-Kamiokande experiment, will be
\begin{itemize}
\item
 Upgrade the J-PARC main ring to the power of 1.3 MW. This would allow to collect $20 \times 10^{21}$ POT by 2026, 3 times more the original beam request of $7.7 \times 10^{21}$ POT and more than 6 times the statistics collected so far in neutrino+antineutrino modes.
\item
 Push the magnetic horns system to its ultimate performances, bringing its operation current from 250 to 320 kA. This upgrade would provide a 10\% increment of neutrino fluxes reducing the wrong-sign contamination by 5-10\%
\item
Upgrade the ND280 close detector as proposed in \cite{Abe:2019whr}. A new tracker, consisting of a 2 ton horizontal plastic scintillator target  sandwiched between two new horizontal TPCs, will substitute the present P$\emptyset$D.
The plastic scintillator, called Super-FGD \cite{Sgalaberna:2017khy}, consists of a matrix of $1\times 1\times 1$ cm$^3$ cubes made of extruded plastic scintillator, where each cube is crossed by three wavelength shifting fibers along the three directions, allowing a 3D reconstruction of the events.

The TPC's will be similar in design to the existing ones, with two major improvements: the Micromegas detector will be constructed with the "resistive bulk" technique,
thereby allowing a lower density of readout pads and eliminating the
discharges (sparks).
 The field cage will be realized with a layer of solid insulator laminated on a composite
material. This will minimize the dead space and maximize the tracking volume.

 This tracker would be surrounded by Time-of-Flight counters (made of
plastic scintillator) to measure the direction of the tracks and reject out of fiducial volume events.
The total mass of active target for neutrino interactions increases from 2.2 
to 4.3 tons, allowing the doubling of the expected statistics for a
given exposure. Furthermore the new configuration will allow a better coverage of the neutrino quasi-elastic interactions, improving the acceptance for high angle tracks.
\end{itemize}

\section{\nova  in the NuMI beam}
\label{sec:nova}

The Neutrinos at the Main Injector (NuMI) neutrino beam \cite{Adamson:2015dkw} is currently
the most powerful neutrino beam in the world and is running with a peak hourly-averaged power of 742 kW. It has been the workhorse of
accelerator neutrino physics in the US since 2005 and served several
experiments both at Fermilab (MINER$\nu$A, ArgoNeuT) and at far
locations (MINOS, MINOS+ in the Soudan mine and \nova on surface at
Ash River, MN). In this review we will mostly focus on its performance and
results as neutrino beam for \nova, which commenced data taking on
February 2014 \cite{Ayres:2007tu}. 

The NUMI beam is produced by the 120~GeV protons extracted from the
Fermilab Main Injector. Like the CNGS, it thus leverages the
acceleration chain of proton colliders steering the protons toward a
solid state target.  

The protons from the Main Injector hit the graphite target and the produced hadrons
are focused by two magnetic horns and then enter a 675~m long decay
volume. The maximum proton energy that can be achieved at the Main
Injector also offers a greater flexibility in the choice of the mean
neutrino energy. This feature was successfully exploited during the
MINOS data taking \cite{Michael:2008bc,Adamson:2014vgd}, when the uncertainty on \dmtt was still quite large
and the location of the first oscillation maximum was 
unclear. After 2014, however, NUMI was operated mostly in ``medium
energy'' mode, producing neutrinos with a mean energy of about 2
GeV. This operation mode maximizes the number of oscillated neutrinos
in \nova, which is located off-axis (14~mrad) 810~km far from Fermilab.

Compared with the NUMI run at the time of MINOS, the current facility exploits some
modification of the acceleration complex that originates from the
shutdown of Tevatron. The Fermilab Recycler is now used extensively
for neutrino physics and stores the Booster cycles during acceleration
in the Main Injector. In this way, the two most time-consuming
operations of NUMI (the transfer from the Booster to the next
accelerator, and the acceleration in the Main Injector) can be carried
out in parallel rather than in series. As a consequence, the overall
cycle time is reduced from 2 to 1.3~s. From February 2014 to June
2019, \nova accumulated $8.85 \times 10^{20}$ pot in neutrino mode and
$12.3 \times 10^{20}$ pot in anti-neutrino mode, reversing the
currents in the focusing horns. The purity of the \numu (\nubarmu)
beam is 96\% (83\%), respectively.

\begin{figure}
\centering
\includegraphics[width=0.8\textwidth]{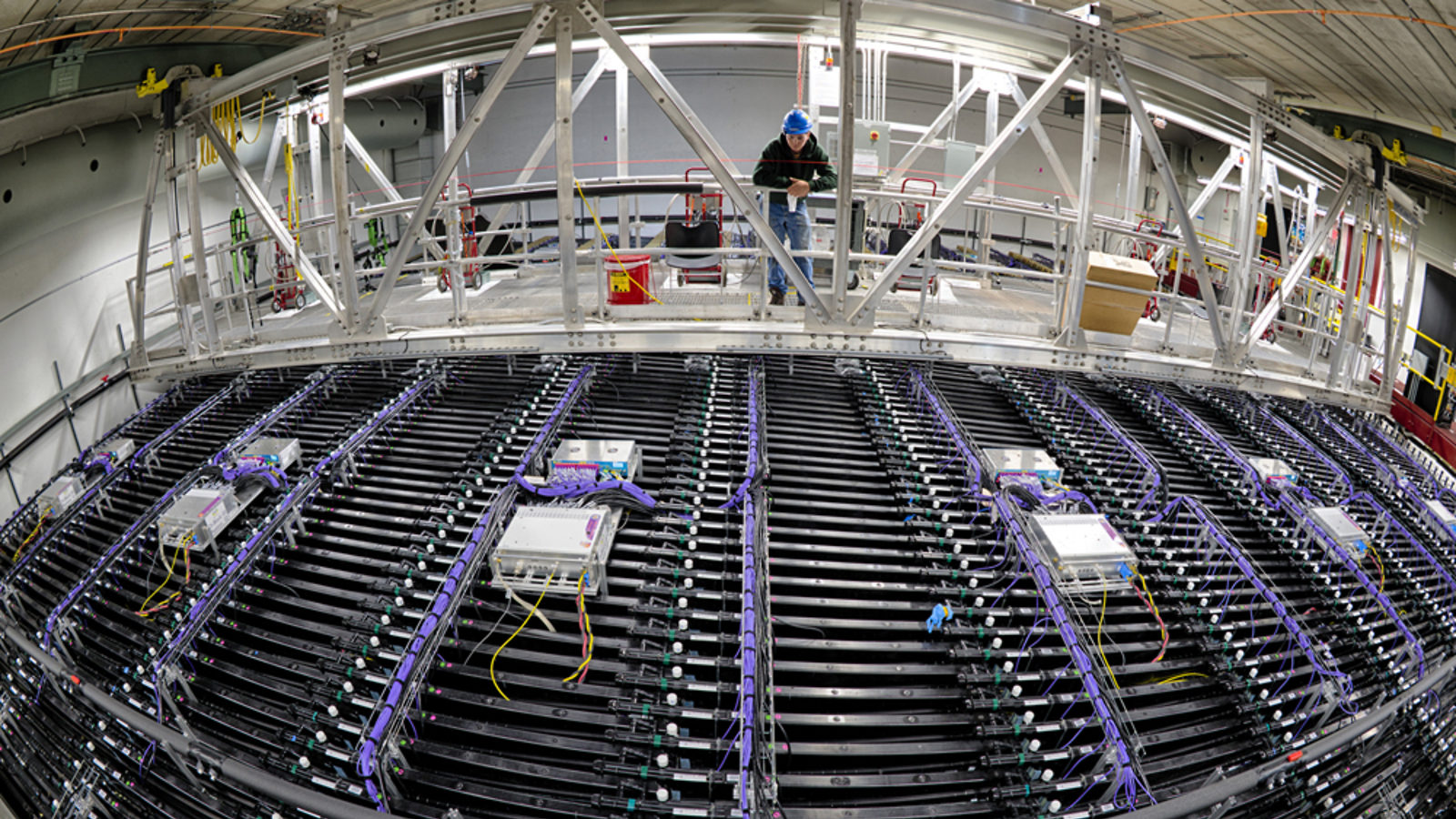}
\caption{The \nova detector at Ash River.}
\label{fig:nova_detector}
\end{figure}

The NOvA experiment measures the energy spectra of neutrino
interactions in two detectors. The Near Detector is placed in the
proximity of NUMI at a distance of 1 km from the production target and
is located 100~m underground.  It is a 290 t liquid scintillator
detector that measures 3.8~m $\times$ 3.8~m $\times$ 12.8~m and is
followed by a muon range stack.  The 14 kton far detector (see Fig.~\ref{fig:nova_detector}) is the
largest neutrino detector ever operated on surface and measures 15~m
$\times$ 15~m $\times$ 60~m. Both detectors use liquid scintillator
contained in PVC cells that are 6.6 cm $\times$ 3.9~cm (0.15 radiation
lengths per cell) in cross section and span the height and width of
the detectors in planes of alternating vertical and horizontal
orientation. The light is read out by wavelength-shifting fibers
connected to APDs. Operation at surface, with practically no
overburden (3 meter of water equivalent, 130 kHz of cosmic-ray
activity) is possible thank to the short duration of the proton burst
that allows to exploit the time correlation between the events observed in the
detector and the occurrence of the proton extraction to the target in NUMI. The analysis~\cite{NOvA:2018gge} is built upon the set of all APD signals above threshold around the 10$\mu$s beam spill. The cosmic ray background is directly measured by a dedicated dataset recorded in the 420~$\mu$s surrounding the beam spill. This control sample is employed for calibration, study of stability response and for the training of the event reconstruction algorithms.

\nova was designed to search for non-null values of \thetaot and
measure the leading oscillation parameters at the atmospheric scale
(\dmtt, \thetatt). Even if the discovery of \thetaot has been achieved
by Daya Bay, Double-Chooz, RENO and T2K well before the start of the NOVA data
taking, the physics reach of this experiment has been significantly
boosted by the large value of this angle. NOVA is the only long baseline experiment currently exploiting
matter effects to determine the neutrino mass pattern. As mentioned in Sec.~\ref{sec:oscillations},
the presence of a dense medium can perturb the oscillation probability of neutrinos because matter is rich in electrons
and deprived of muons and tau leptons. As a consequence, the evolution of the flavor wavefunction is perturbed by the fact that
\numu and \nutau can interact only by neutral currents (NC) with the electrons of matter while both NC and CC elastic scattering can occur for \nue.
Elastic weak scattering can occur also in the nuclei but in this case the \nue have the same interaction probabilities as for \numu or \nutau. The matter perturbations thus are only due to the density of the electrons in the medium and are parameterized by an effective potential $V = \sqrt{2} G_F n_e(x)$. Here, $G_F$ is the Fermi constant and $n_e(x)$ is the electron matter density at the point $x$. Unlike propagation in the sun, where $n_e(x)$ is strongly varying and can cause highly non linear phenomena (MSW resonant flavor conversions \cite{giunti-kim}), the electron density in the earth crust and mantle is nearly constant.  The oscillation formula corrected by matter effects in conditions interesting for
long baseline experiment can be written as \cite{Cervera:2000kp,Freund:2001pn}
\bea
  P (\nu_\mu \to \nu_e) &\! \simeq &\!
    \sin^2 \theta_{23}\, \frac{\sin^2 2 \theta_{13}}{(A-1)^2}\, \sin^2 \left[ (A-1) \Delta_{31} \right]
    +\, \alpha^2 \cos^2 \theta_{23}\, \frac{\sin^2 2 \theta_{12}}{A^2}\, \sin^2 (A \Delta_{31})  \nn \\
  &\! &\! +\ \alpha\, \frac{\cos \theta_{13} \sin 2 \theta_{12} \sin 2 \theta_{13} \sin 2 \theta_{23} \cos \delta_{CP}}{A (1-A)}\,
    \cos \Delta_{31} \sin (A \Delta_{31}) \sin \left[ (1-A) \Delta_{31} \right]  \nn \\
  &\! &\! -\ \alpha\, \frac{\cos \theta_{13} \sin 2 \theta_{12} \sin 2 \theta_{13} \sin 2 \theta_{23} \sin \delta_{\rm CP}}{A (1-A)}\,
    \sin \Delta_{31} \sin (A \Delta_{31}) \sin \left[ (1-A) \Delta_{31} \right]\, ,  \hskip 1cm
\label{eq:numu_nue_3f_matter}
\eea
where $\alpha \equiv \Delta m^2_{21} / \Delta m^2_{31}$, $\Delta_{31} \equiv \Delta m^2_{31} L / 4 E$
and the electron density is embedded in 
$A \equiv 2 V E / \Delta m^2_{31} = 2 \sqrt 2\, G_F n_e E/ \Delta m^2_{31}$.
The probability for $\bar \nu_\mu \to \bar \nu_e$ oscillations is the same as for neutrinos except for a change of
the sign of the CP-violating phase $\delta_{\rm CP}$ and of the matter parameter $A$.
Eq.~(\ref{eq:numu_nue_3f_matter})
is an approximate formula that results from expanding to second order in the small quantities $\alpha$ and $\sin \theta_{13}$ under the hypothesis of a constant electron density $n_e$. Exact formulas can be obtained integrating the propagation in steps of $x$ and are commonly embedded in software as GLoBES \cite{Huber:2004ka}. Still, Eq.~(\ref{eq:numu_nue_3f_matter}) is an excellent approximation for all current long-baseline experiments. The change of sign of $A$ is hence a fake CP violation effect due to the presence of electrons in matter and should be decoupled by genuine CP effects. On the other hand, fake CP effects depend on the sign of \dmtt and, therefore, provide an elegant tool to determine
whether $m_1 < m_2 < m_3$ (``normal ordering'', NO) or $m_3 < m_1 < m_3$ (``inverse ordering'', IO). All other ordering options are already excluded by solar neutrino data that imply $\dmot >0$. Distinguishing $\dmtt<0$ (IO) from $\dmtt>0$ (NO) is extremely interesting not only for building theory models of neutrino masses and mixing. IO would imply that the eigenstate that is maximally mixed with $\nu_e$ is indeed the largest one. IO thus eases remarkably the measurement of the absolute neutrino masses and of neutrinoless double beta decay,  which are entirely based on $\nu_e$. NOVA is the only running experiment sensitive to matter effects because $A=0.046$ in T2K and $A=0.12$ in \nova . Similarly, the CNGS is not sensitive to matter effect because it is off the peak of the oscillation maximum ($\Delta_{31} \ll 1$) \cite{Migliozzi:2003pw} and hence,
$$
\frac{ \sin^2 (1-A)\Delta_{31} }{(A-1)^2} \simeq \Delta_{31} 
$$
In the design phase of \nova, it was not clear whether it would had
been possible (even with a combination with T2K) to disentangle mass
ordering effects from genuine CP effects because accidental
cancellations \cite{ambiguity1, ambiguity2, ambiguity3} arise for some values of $\delta$ and sign(\dmtt)
(``$\delta-sign(\dmtt)$ ambiguity'').  Matter effects enhance the
electron neutrino appearance probability in the case of normal mass
hierarchy (NO) and suppress it for inverted mass hierarchy (IO).
CP-conserving oscillations occur if $\delta = 0$ or $\pi$, while the
rate of \nue CC is enhanced around $\delta = 3\pi /2$, and suppressed
around $\delta = \pi /2$. At \nova the impact of these factors on the
\nue appearance probability are of similar magnitudes, which can lead
to degeneracies between them, especially in neutrino runs alone. For
antineutrinos, the mass hierarchy and CP phase have the opposite
effect on the oscillation probability. Increasing values of
$\sin^2 \thetatt$ increase the appearance probabilities for \nue and
\nubare alike.  Still, the current best fit of $\delta$ are in the
most favorable region of the parameter space and, if the central
values are confirmed, \nova will be able to gain a strong evidence of
NO in the next few years  combining neutrino and antineutrino runs \cite{nova_neutrino2018}.

\nova is a detector with a very low density so that electromagnetic
showers develop for several PVC cells easing the identification of
\nue CC events. In addition, \nova is equipped with a near detector,
which opens up the possibility of measuring neutrinos  at the far detector also in disappearance mode ($\nu_\mu \rightarrow \nu_\mu$). The energy of the beam is well below the kinematic threshold for $\tau$
production and therefore the contamination from $\nutau$ CC events is
negligible.  

As for T2K, the simulation of \nova employs a detailed description of the beamline based on GEANT4. Again, the computed neutrino 
flux is corrected according to constraints on the
hadron spectrum from thin-target hadroproduction data
using the PPFX tools developed for the NuMI beam by
the MINERvA collaboration~\cite{Aliaga:2016oaz}.  
Neutrino interactions in the detector are simulated with GENIE~\cite{genie} while the detector response is simulated with GEANT4. Unlike T2K, \nova employs in a systematic manner analyses
techniques inherited from the field of computer vision.  The detector
hits are formed into clusters by grouping hits in time and space to
isolate individual interactions. A Convolutional Visual Network
classifier employs the hits from these clusters, without any further
reconstruction, as input and applies a series of trained operations to
extract the classifying features from the image \cite{Acero:2019ksn}. 
The classifier feeds an artificial neutral network, i.e. an input layer of perceptrons linked to additional hidden layers and an output node. The network is then trained
to recognize different neutrino topologies: $\nu_e$ CC, $\nu_\mu$ CC
and NC. The $\nu_\mu$ CC and $\nu_e$ CC samples correspond to events whose score exceed a given threshold. Muon identification in $\nu_\mu$ CC events  is performed employing the $dE/dX$ and multiple scattering of the track, the length of the track and the fraction of plane where the track is mip-like~\cite{NOvA:2018gge}. The \nova analysis selects $\nu_\mu$ CC events with an efficiency of 32.2\%. In terms of purity, the final selected sample is 92.7\%  $\nu_\mu$ CC. The energy of the muon is estimated from the range and the hadronic energy from the number of hits. The energy resolution for the whole sample of $\nu_\mu$ CC is 9.1\%. The quality of the simulation and event selection procedures are validated using the $\nu_\mu$ CC reconstructed spectrum at the near detector, which is not distorted by oscillations. Possible discrepancies (in the 4-13\% range for $\nu_\mu$ CC) are corrected to reach complete agreement with the near detector (ND) data.  
The \dmtt and $\sin^2 2\theta_{23}$ parameters are extracted
from the spectrum of the $\nu_\mu$ CC events at the far detector compared with the corresponding spectrum at the near detector. 
The selection of the $\nu_e$ CC sample is particularly difficult due to the presence of electromagnetic components in NC events (e.g. $\pi^0$ production) or short muon in $\nu_\mu$ CC events. The near detector $\nu_e$ CC selected sample consists indeed of 42\% $\nu_e$ CC, 30\% NC and 28\% $\nu_`\mu$ CC. The selection efficiency -- i.e. the probability to select a true $\nu_e$ CC event -- amounts to 67.4\%. The estimator of the energy of the electron candidate has a resolution of 11\%. Since the NC and cosmogenic background is mostly in the low energy range of the electron spectrum and the $\nu_e$ CC from the $\nu_e$ contamination of the beam dominates at large energies, the analysis is performed only for events whose energy is between 1 and 4 GeV.
Again, the quality of the simulation is validated by ND data. In particular, $\nu_\mu$ selected candidates are used to cross-check the $\nu_e$ selection efficiency and correct the expected $\nu_e$ signal at the far detector (FD). $\nu_e$ at ND are artificially created in the simulation replacing the muon with an electron of the same energy and direction. The $\nu_e$ CC selection algorithm is applied to this sample and the results match expectations within 2\%. The intrinsic \nue
component of the beam (0.7\% of all neutrinos) represents a nearly
irreducible background.
ND data are used to constrain such $\nu_e$ component beyond the estimate performed with GEANT4. ND data are also used to constrain the observed $\nu_\mu$ CC and NC events (see Fig. 5 of Ref.~\cite{NOvA:2018gge}). For the 2018 \nova analysis, corresponding to $8.85 \times 10^{20}$ pot, the overall error budget for $\sin^2 \theta_{23}$, \dmtt\  and $\delta$ are $4 \times 10^{-2}$, $9 \times 10^{-5}$ eV$^2$ and 0.67$\pi$, respectively. At present, the systematic budget is smaller than the statistical uncertainty and amounts to  $1 \times 10^{-2}$, $4 \times 10^{-5}$ eV$^2$ and 0.12$\pi$ for $\sin^2 \theta_{23}$, \dmtt\  and $\delta$, respectively. However, it is likely that  \nova will be systematic limited in the forthcoming years, well before the start of DUNE.

\begin{figure}[H]
\centering
\includegraphics[width=0.8\textwidth]{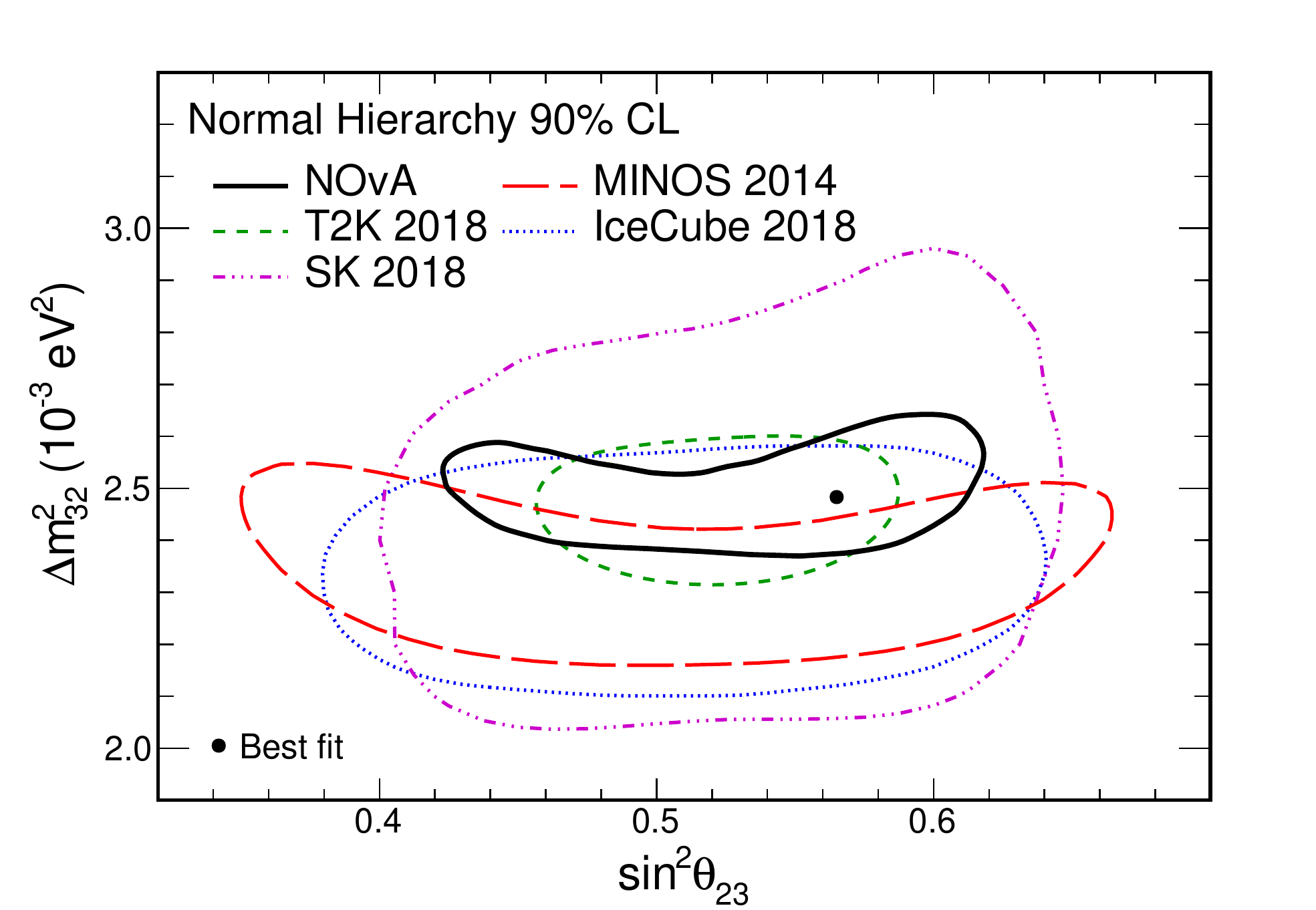}
\caption{\dmtt versus $\sin^2 \theta_{23}$ (90\% CL) measured by \nova, T2K, MINOS+ and  atmospheric neutrino experiments.}
\label{fig:nova_fit}
\end{figure}

At the time of writing~\cite{nova_win2019}, the T2K and \nova results \cite{Adamson:2016xxw,Adamson:2017gxd,Adamson:2017qqn,NOvA:2018gge} are compatible for all oscillation parameters (see Fig.~\ref{fig:nova_fit}).
\nova data indicate a slight preference (1.8 $\sigma$) for normal hierarchy ($\dmtt>0$), which corroborates the hint from
atmospheric neutrino data \cite{Esteban:2018azc}.

\section{DUNE and Hyper-Kamiokande}
\label{sec:dune}

The optimal strategy to perform precision measurements and three family interference effects in the neutrino sector has been debated for decades. Accelerator neutrino beams became the technology of choice already in late 90s. At that time, experimental data start pointing
toward two different mass scales (the atmospheric mass scale \dmtt and the solar mass scale \dmot) but the ratio of the scales ($\alpha \simeq 0.03$ in eq.~\ref{eq:numu_nue_3f_matter}) was not too small to prevent the observation of perturbations driven by \dmot in the leading
\numunue transitions at the atmospheric scale. As a consequence, an intense and controlled \numu source of artificial neutrinos at the O(GeV) energy scale would have been the optimal facility to search, for instance, for CP violation in the leptonic sector. Unfortunately, at that time, \thetaot was not know (the Chooz limits indicated $\thetaot < 12^\circ$) and the overall size of the leading \numunue oscillation could have been arbitrarily small. For $\thetaot \ll 12^\circ$, the intensity of conventional beam would have been prohibitive at baselines O(1000~km) and the intrinsic contamination would have represented a major limitation to seek for \numunue oscillations at the per-mill level or below \cite{Mezzetto:2010zi, Battiston:2009ux}. This consideration triggered an intense R\&D toward new acceleration concepts, namely the Neutrino Factory (see Sec. \ref{sec:future}) and the Beta Beam \cite{Zucchelli:2002sa,mezzetto_beta_beams} that aimed at providing pure and intense sources of \nue and \numu. The discovery of \thetaot at $8^\circ$ set the scale of the leading \numunue oscillation at the level of 7\% ($\sin^2 2\thetaot \simeq 0.07$), which can be studied  further increasing the power of current accelerator beams. Since 2012, therefore, R\&D efforts have focused toward the design of ``Superbeams'': conventional neutrino beams with beam power exceeding 1 MW. 

Two Superbeams will drive the field of accelerator neutrino physics in the  next twenty years: the Long Baseline Neutrino Facility (LBNF) at Fermilab serving the DUNE experiment in South Dakota and the J-PARC Neutrino Beamline upgrade serving Hyper-Kamiokande in Japan.
Both facilities are able to address CP violation in a significant region of the $\delta$ parameter space, perform precision measurements
of oscillation parameters (e.g. the octant of \thetatt, which is still unknown), determine the mass hierarchy using matter effect in beam (DUNE) or atmospheric (Hyper-Kamiokande) neutrinos and, in general, establish stringent tests for the minimally extended SM and its competitors. Furthermore, both DUNE and Hyper-Kamiokande are sensitive to natural neutrino sources and can perform a rich astroparticle physics programme.  
\subsection{DUNE}
LBNF \cite{Acciarri:2016crz} is a new neutrino facility built upon the Fermilab acceleration complex, as it was in the past for NUMI. In the next few years this complex will undergo a significant upgrade (Proton-Improvement-Plan PIP-II) that includes the construction and operation of a new 800 MeV superconducting linear accelerator. As for NUMI, the primary proton beam will be extracted from the
Main Injector in the energy range of 60-120 GeV. With the Main Injector upgrades already implemented for
\nova as well as with the expected implementation of PIP-II, each extraction will deliver
$7.5 \times 10^{13}$ protons in one machine cycle (0.7~sec/60~GeV - 1.2~sec/120~GeV) to the LBNF target in 10~$\mu$s. The complex
delivers to DUNE an average power of 1.2 MW for 120 GeV protons. Neutrinos
are produced after the protons hit a solid target and produce mesons which are subsequently
focused by three magnetic horns into a 194 m long helium-filled decay pipe where they decay into muons
and neutrinos. The facility is equipped with a near detector located 300 m after the absorber. Unlike its predecessor, LBNF has been designed to provide neutrinos over a wide energy range in order to cover both the first and the second oscillation maximum in DUNE. 
The beamline is designed to cope with a 2.4 MW power beam without retrofitting since an additional upgrade of the proton acceleration chain is planned (PIP-III) to be operational by 2030.

\begin{figure}
\centering
\includegraphics[width=0.8\textwidth]{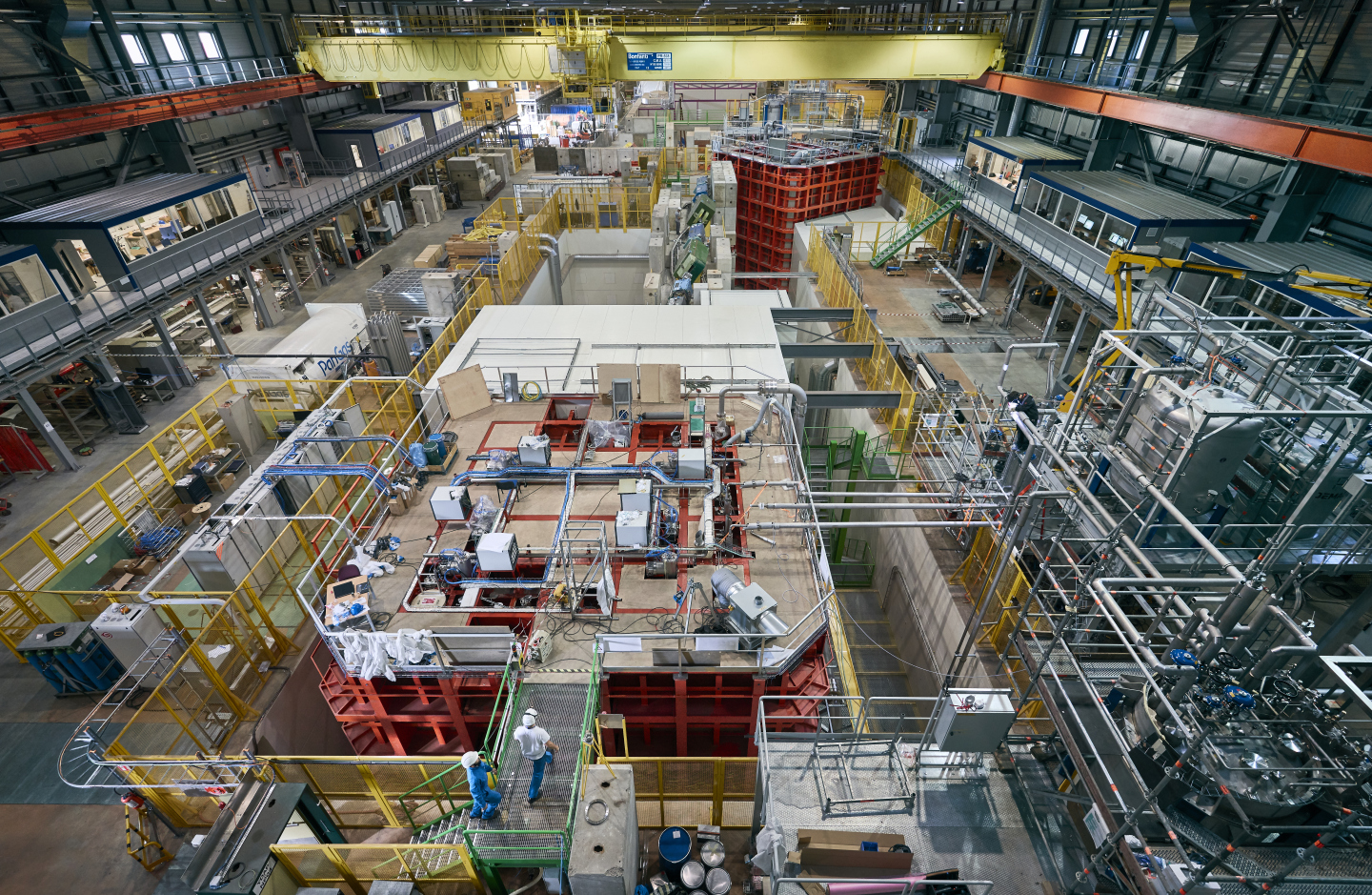}
\caption{The ProtoDUNE-SP and ProtoDUNE-DP detectors at CERN.}
\label{fig:protodune}
\end{figure}

The 40 kt DUNE Far Detector \cite{Abi:2018alz} consists of four LArTPC detector modules,
each with fiducial mass of about 10 kt, installed approximately 1.5~km
underground and on-axis with respect to the LBNF beam.  Each of the LArTPCs fits inside a cryostat of internal
dimensions 18.9~m (width) $\times$ 17.8~m (height) $\times$ 65.8~m
(length) that contains a total Argon mass of about 17.5 kt. The four
identically sized modules provide flexibility for staging construction
and for evolution of LArTPC technology. The baseline design has been
recently validated at CERN by means of a 770 ton prototype
(ProtoDUNE-SP \cite{Abi:2017aow}, with a fiducial mass of about 400 ton) using charged particle beams and cosmic rays. This
``single phase'' (SP) designs inherits the most important techniques
developed by ICARUS (see Sec.~\ref{sec:cngs}) for the charge readout, HV and the Argon purification system. Unlike ICARUS,
ProtoDUNE-SP (see Fig.~\ref{fig:protodune}) is engineered to be scalable up to  a DUNE
module. It means that the components of the TPC are modular and can be
assembled on-site in the underground laboratory and the cryogenic
system has been greatly simplified. ProtoDUNE-SP demonstrated that it
is possible to achieve an electron lifetime comparable with ICARUS using
a cryostat based on a very cheap technology  developed
for industrial applications. A ``membrane cryostat'' is made of a
corrugated membrane that contains the liquid and gaseous Argon, a
passive insulation that reduces the heat leak, and a reinforced
concrete structure to which the pressure is transferred. A secondary
barrier system embedded in the insulation protects it from potential
spills of liquid Argon, and a vapor barrier over the concrete protects
the insulation from the moisture of the concrete. This system is used
since long for the transportation of liquified natural gases and has been
tailored for use with high purity Argon during the R\&D for DUNE.
Unlike ICARUS, the charge readout electronic is located inside the
cold volume to reduce noise and the light detection system is based on
Silicon Photomultipliers (SiPMs) embedded in the anode wire planes instead of large area PMTs. 

For the second module, DUNE is considering an innovative Double Phase
LArTPC where the electrons are multiplied in the gas phase. The Double
Phase (DP) readout allows for lower energy threshold and a reduction
of complexity and cost due to the longer drift length: in DP the
distance between the cathode and the anode is 6 m and the drift takes place
in the vertical direction while the distance of the horizontal planes
in a SP module is 3.5~m.  This technique is currently under validation at CERN
(ProtoDUNE-DP \cite{Agostino:2014qoa}).

Both DUNE and Hyper-Kamiokande has been designed firstly to provide a
superior sensitivity to CP violation and mass hierarchy compared with
the previous generation of long-baseline experiments.  DUNE, in
particular, is able to establish mass ordering at $5 \sigma$ in 2.5
years of data taking whatever is the value of $\delta$ \cite{Abi:2018dnh}. Similarly, it
can establish CP violation in the leptonic sector at 5 (3) $\sigma$
level in 10 (13) years for 50\% (75\%) of all possible values of
$\delta$.  If the value of $\delta$ is large, as suggested by current
hints, CP violation can be established in about 6 years of data taking:
statistics and control of the systematics will thus be crucial to
improve the precision in the measurement of $\delta$ (about $20^\circ$
for $\delta=-\pi/2$). Table~\ref{tab:milestones} summarizes the sensitivity for a selection of physics observables as a function of time. The number of years are computed assuming the standard DUNE  staging scenario: the start of data taking of DUNE will commence after the commissioning of two DUNE modules (20 kton) with an average beam power of 1.2 MW (T0). In the next year (T0+1y), the third module will be in data taking (30 kton). DUNE will reach its full mass (40 kton) at T0+3y. The beam will be upgraded to 2.4 MW three years later, i.e. at T0+6y.

\begin{table}
\caption{Time to achieve the most relevant physics milestones in DUNE for the current detector staging scenario (see text).}
\centering
\begin{tabular}{cc}
\toprule
Milestone	& Exposure (years)  \\ \hline
5$\sigma$ mass hierarchy at $\delta=-\pi/2$  & 1 \\ 
5$\sigma$ mass hierarchy at any $\delta$   &  2.5 \\
3$\sigma$ CP violation at $\delta=-\pi/2$ &   3 \\
3$\sigma$ CP violation at 50\% $\delta$ coverage &   5 \\
5$\sigma$ CP violation at $\delta=-\pi/2$ &   7 \\
5$\sigma$ CP violation at  50\% $\delta$ coverage &   10 \\
3$\sigma$ CP violation at  75\% $\delta$ coverage &   13 \\
$\delta$ resolution of 10$^\circ$ for $\delta=0$ & 8 \\
\hline 
\end{tabular}
\label{tab:milestones}
\end{table}

As noted above, this class of experiments is able to test in a single
setup the consistency of the minimally extended SM employing the rate
of \numu and \nue in neutrino and antineutrino runs. It thus has a
rich search program of physics beyond the Standard Model. Natural sources
are exploited in DUNE for the study of Supernovae Neutrino Bursts
(SNB), atmospheric neutrinos and the search for proton decay. In
particular, DUNE is the only large mass experiment based on
Argon. It has a particular sensitivity to the $\nu_e$ component
of the SNB through the
$\nue {\rm ^{40}Ar} \rightarrow e^- \ {\rm ^{40}K}^*$ reaction and to proton decay in
the $p \rightarrow K^+ \overline{\nu}$ channel where the kaon is below the kinematic threshold in Cherenkov detectors.

\vspace{0.5cm} \noindent
\subsection{Hyper-Kamiokande}
The possibility of a 5$\sigma$ sensitivity on the measurement of leptonic CP violation, the very reach program of astrophysical measurements, low energy neutrino multimessenger physics and proton decay, and the strength of the double Noble prize (Super)Kamiokande experiment, motivated a large international collaboration to propose the Hyper-Kamiokande (Hyper-K) project \cite{Abe:2018uyc}.

\begin{figure}
\centering
\includegraphics[width=0.7\textwidth]{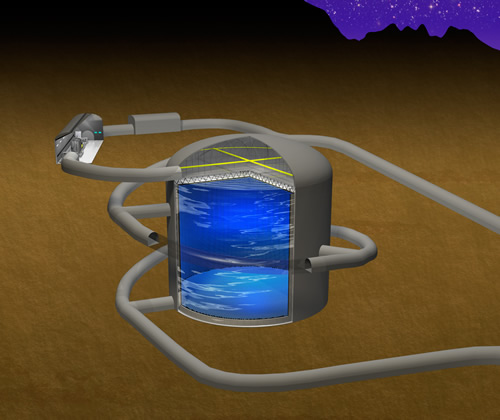}
\caption{Sketch of the Hyper-Kamiokande detector.}
\label{fig:HK}
\end{figure}

The design of Hyper-Kamiokande is a cylindrical tank with a diameter of 74 m and height of 60 m.
 The total (fiducial) mass of the detector is 258 (187) kt,
 for a fiducial mass 8 times larger than Super-Kamiokande.
A schematic view of Hyper-Kamiokande is shown in Fig.\ref{fig:HK} 

 The Hyper-Kamiokande detector will be located in Tochibora mine, 8 km south of
Super-Kamiokande but at the same distance from J-PARC (295 km) and at the same off-axis angle ($2.5^\circ$). With an overburden of 1750 meters-water-equivalent the muon flux is about  5 times higher than
in Super-Kamiokande. 
Hyper-Kamiokande will be instrumented with 40,000  inward facing PTMSs, the 50 cm diameter Hamamatsu R12860 PMTs which have ~2 times higher
photon detection efficiency than those of  Super-Kamiokande, for a photocoverage fraction of 40\%. 

Another possibility is to have a photocoverage fraction of 20\%
 complemented by 5000 Multi-PMT derived from the design of KM3NeT \cite{Vivolo:2018grv}.
Within a 50 cm  diameter they will contain 19 7.7-cm PMTs together with integrated readout and calibration \cite{Ruggeri:2018vnc}. They can allow an increased granularity, enhancing event reconstruction in particular for multi-ring events, and better timing, which could further reduce the dark hit background and event reconstruction. The improved granularity should also guarantee an increase of the fiducial volume.
Simulations show that this second setting-up guarantees very similar performances of the original 40\% coverage as far as concerns beam oscillation studies.

The outer detector veto will be similar to that of SK, with 20 cm PMTs resulting in a 1\%  photocathode converage of the inner wall.

The neutrino beam configuration of Hyper-K would be the same as T2K, with an upgraded power of the main 30 GeV proton ring at 1.3 MW.

The close detector system is designed to be again ND280, in the upgraded version described in Section \ref{sec:T2K2}, 

A water Cherenkov (WC) near detector can be used to measure the cross section on $\rm{H_2O}$ directly.
It could be placed at intermediate distance, $\sim$ 2~km, from the interaction target, in order to have a neutrino flux more similar to the flux at SK and an interaction rate not critical for the performance of a WC detector.
These additional WC measurements are important to achieve the low systematic errors required by Hyper-K, and  complements those of the ND280 magnetised tracking detector.

\begin{figure}
\centering
\includegraphics[width=0.8\textwidth]{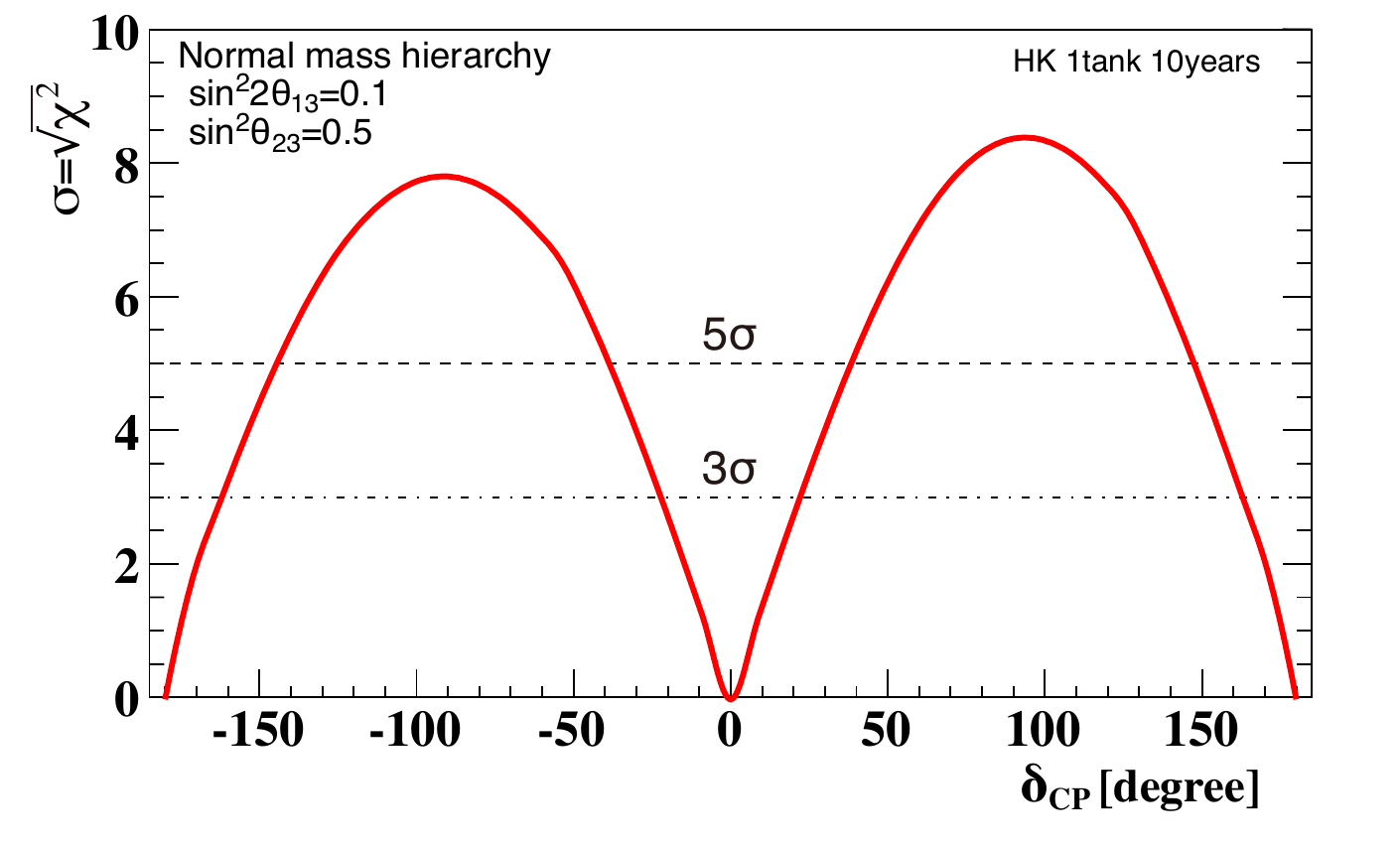}
\caption{Sensitivity of Hyper-K in measuring $\delta_{CP}$ \cite{Abe:2018uyc}.}
\label{fig:CP-HK}
\end{figure}

The studies of the sensitivity to CP violation assume an integrated beam power $1.3 \; \rm{MW} \times 10^8$~s
corresponding to $2.7\times 10^{22}$ protons on target (POT) with the 30 GeV J-PARC proton beam.
The selection criteria of $\nu_e$  and $\nu_\mu$ candidate events are based on those
established in the T2K experiment. The total systematic uncertainties of
the number of expected events are assumed to be 3.2\% for the $\nu_e$ appearance and 3.9\% for the $\overline{\nu}_e$
appearance, assuming a 1:3 ratio of beam power for the $\nu$:$\overline{\nu}$ modes. These systematic error values roughly correspond to the statistical errors at full statistics.
 Fig.\ref{fig:CP-HK}  shows the expected significance to exclude the CP conserving cases at full statistics.
 CP violation in neutrino oscillations
can be observed with $> 5(3)\sigma$ significance for 57(80)\% of the possible values of $\delta_{CP}$. 
Exclusion of $\delta_{CP}=0$ can be obtained with a significance of 8$\sigma$ in the case of maximal CP violation with
$\delta_{CP} = -\pi/2$, the favored value of T2K.
By combining the analysis of beam neutrinos with the analysis of atmospheric neutrinos, Hyper-K would have a $\sim 5\sigma$ sensitivity in rejecting the false value of neutrino mass ordering.

A powerful possible upgrade of Hyper-K would consist in building a second identical detector in Korea at a distance of $\sim 1100$ km and larger overburden ($\sim 2700$ mwe) from J-PARC  \cite{Abe:2016ero}. This second far detector can be exposed by 1-3 degree off-axis neutrino beams from J-PARC and
would be then mostly sensitive to neutrino oscillations at the second oscillation maximum.

 This two-baseline setup could  break the degeneracy of
oscillation parameters and achieve a better precision in measuring $\delta_{CP}$.

Among the non oscillation physics goals of Hyper-K, is worth to note the 90\% sensitivity to proton decays of $7.8 \times 10^{34}$ years for the $p \rightarrow e^+ \pi^\circ$ decay channel (existing limit is $1.6 \times 10^{34}$ years) and $3.2 \times 10^{34}$ years for the $p \rightarrow \mu^+ \pi^\circ$ channel (existing limit: $0.7 \times 10^{34}$ years).
Hyper-K would collect $(50-80) \times 10^3$ events for a 10
kpc supernova explosion and $(2-3) \times 10^3$ events for a supernova  at the Large Magellanic Cloud where SN1987a was located.
It could achieve a $4.2\,\sigma$ statistical sensitivity in detecting supernova relic neutrinos (SRN), integrating 10 years of data taking.
These numbers are computed not assuming gadolinium doping in the detector, that would be highly beneficial in all these channels \cite{Watanabe:2008ru}.
In particlular with 0.1\% by mass of gadolinium dissolved in water, the threshold for SRN detection would decrease from 16 to 10 MeV, allowing to explore the history of supernova bursts back to the epoch of red shift $(z)\sim 1$. The number of detected SRN events with gadolinium loading
would become about 280, compared with 70 events without loading.

\section{The systematic reduction programme}
\label{sec:sys}

Neutrino oscillation experiments mostly rely on the near-far detector cancellation to mitigate the effects of systematic errors. This technique is the cornestone of long-baseline experiments and it consists of building an additional detector at a distance from the source where oscillations have not developed, yet. The near detectors thus measure the \numu and \nue interaction rate at source and are compared with the rate at the far detector. If the near detector is close to the source (a few hundreds metres) and identical to the far detector, it provides a normalization for the \numu and \nue rates that can be used to extract \numunue $\ $and \numunumu probabilities accounting for the finite efficiency of the detector for \numu and \nue CC events, provided that the corresponding cross sections are properly known. If the detectors can measure both \numu and \nue and the beam is monochromatic, the remaining systematics after the near-far cancellation is $\sim \sigma_{\nu_e} \epsilon_{\nu_e}/ \sigma_{\nu_\mu} \epsilon_{\nu_\mu}$. Unfortunately, other effects make the near-far detector comparison more challenging if systematics must be kept well below 10\%. Firstly, the beam spectrum observed at the far location is different from the beam at source as observed in the near detector. This is due to the large difference between the angular acceptance of the two detectors. Such effect is exacerbated if the far detector is located off-axis with respect to the beam. In addition, high power beams have such a large rates at near site that pile up effects may jeopardize the efficiency of the detector, increase the background of interactions in the surrounding material and change the performance of PID algorithms. This is particularly disturbing for technologies that are intrinsically slow (liquid Argon TPC) or difficult to be operated at high pile-up rates (Cherenkov detectors). Pile-up effects can be mitigated by using fast neutrino detectors at the price of having the near detector different from the far detector. This is the method of choice of T2K for the ND280 and its upgrades (see Sec. \ref{sec:t2k}).
Finally, since no beam is monochromatic, the incoming neutrino energy must be reconstructed from the kinematics of final state particles, which in turn depend on final state interactions and nuclear effects. The simulation of these effects from first principles is beyond the state of electroweak nuclear physics and the models implemented in the neutrino generators must be tuned and validated by experimental data from either the near detector or dedicated cross section experiments.

Over the years, the most rewarding strategy has been constraining the near detector observables embedding all available information on flux, cross section, energy reconstruction and detector effects combining both measurements on-site (near detector) and data recorded in dedicated facilities (hadroproduction experiments, cross section experiments, detector test-beam). In spite of the complexity of this approach, current long baseline experiments can claim an overall systematic budget at $\sim$ 5\% level for $\nu_\mu$ events and 8-9\% for $\nu_e$ events (see Sec.\ref{sec:t2k}).

\begin{figure}
\centering
\includegraphics[width=0.6\textwidth]{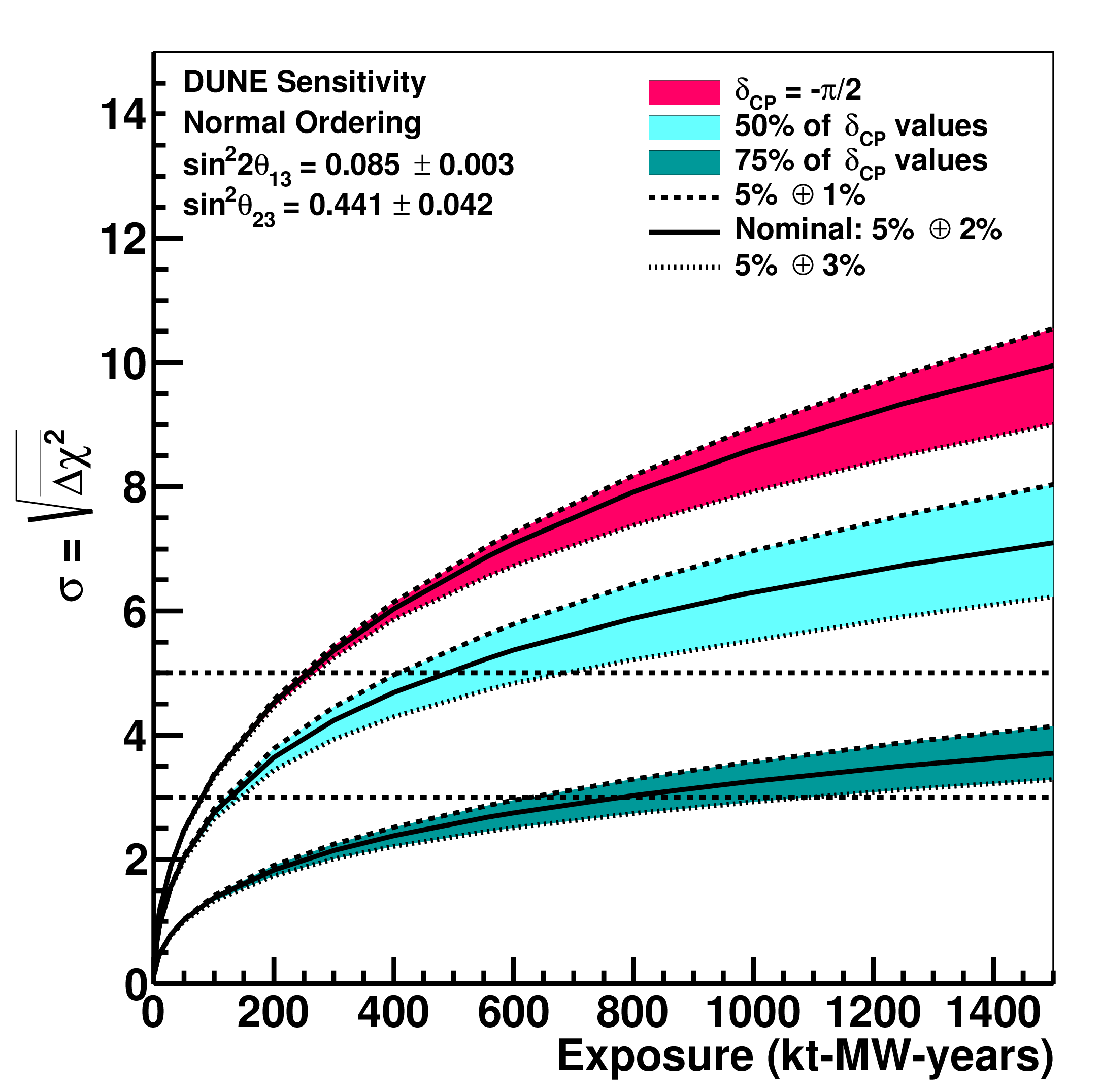}
\caption{CP violation sensitivity for DUNE as a function of the exposure (see text).  The width of the band corresponds to the difference in sensitivity between $\nu_{e}$ signal normalization uncertainty of 1\% and 3\%. The nominal uncertainty of 5\% on the $\nu_{\mu}$ disappearance mode and 2\% on the $\nu_{e}$ appearance mode is shown as the solid line. The three curves show the significance when $\delta_{CP} = - \pi/2$ and the minimum significance for 50\% and 75\% of true $\delta_{CP}$ values (CP coverage).}
\label{fig:effect_systematics}
\end{figure}

The next generation of Super-beams (DUNE and Hyper-Kamiokande) has a physics reach that will be mostly dominated by the residual systematic errors if they remain at the same values of T2K and NO$\nu$A (statistical errors at full statistics will be below 3\%).
Systematic errors have to be roughly halved if the investments for the huge exposures of next generation experiments have to be fully exploited.
For sake of illustration, Fig. \ref{fig:effect_systematics} shows the impact on the $\nu_e$ normalization uncertainty for DUNE as a function of the exposure expressed in (mass)$\times$(beam power)$\times$year, i.e. kton$\times$MW$\times$years~\cite{Abi:2018alz}.

A vigorous systematic reduction programme is therefore extremely cost effective and, in addition, provides robustness against hidden systematic biases in Super-beam experiments. This programme is under development and follows three lines of research.

\subsection{Near detectors}

Both for DUNE and Hyper-Kamiokande, the near detector will comprise a high granularity setup similar to ND280. Even if the technology envisaged in the upgrade of ND280 and in the DUNE high-pressure TPC are different from the far detector, they allow for a superior reconstruction of final state particles in CC events and can be operated at high rate. In DUNE, the high pressure (gas Argon) TPC is complemented by a modular liquid Argon TPC that is readout by pixels after a small (50 cm) drift length \cite{bross_nufact2019}. Unlike the on-axis detector (3DST embedded in the former KLOE magnet and calorimeter), this setup is movable horizontally in the plane perpendicular to the beam axis. It thus samples different beam angles to retrieve the relation between the incident and reconstructed neutrino energies in the detector ("PRISM concept" \cite{Bhadra:2014oma}).
The same concept is under consideration in the Hyper-Kamiokande intermediate detector. This detector is located at a larger distance ($\sim 2$ km) than the N280 to reduce pile-up nuisance. In its nominal position, it samples a beam whose spectrum is the same as for the far detector. The intermediate detector, however, is movable in the vertical axis and thus implements the PRISM concept in the energy range of interest for T2K and Hyper-K \cite{Abe:2018uyc}.

\subsection{Hadroproduction experiments}

\begin{figure}[H]
\centering
\includegraphics[width=0.8\textwidth]{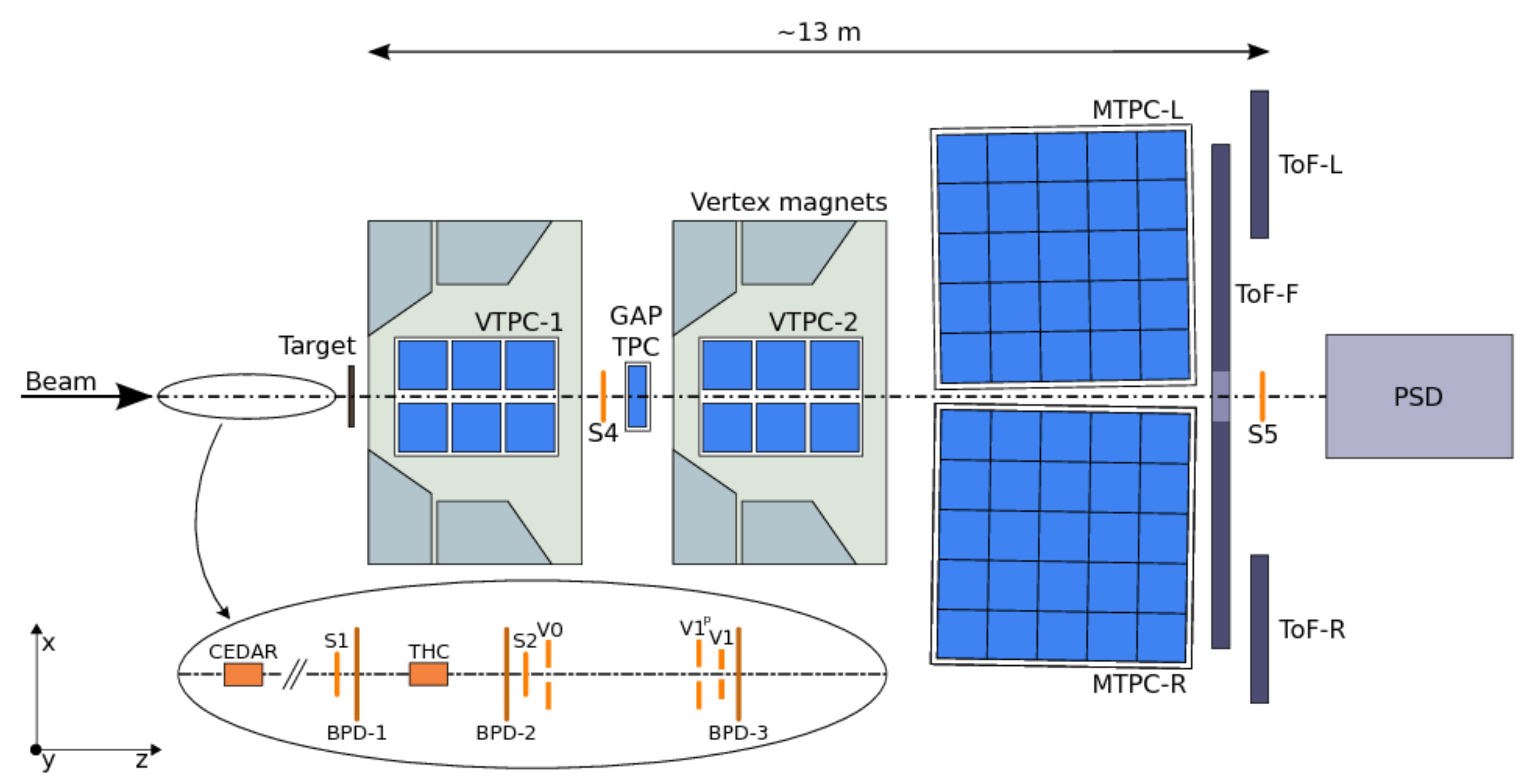}
\caption{Layout of the NA61/SHINE experiment \cite{na61}.}
\label{fig:na61}
\end{figure}

Hadroproduction experiments measure the total and differential yields of secondary particles (mostly pions, protons and kaons) produced by high energy protons impinging on a target. They play a key role in the determination of the initial flux of neutrino beams because the secondary production is dominated by non-perturbative QCD effects and it is thus the most systematic prone part of the beamline simulation. Dedicated hadroproduction experiments have been routinely performed since the 90s as ancillary experiments for long and short baseline facilities and are essential for the DUNE/Hyper-K programme.
Following the experience of the HARP experiment, that measured hadroproduction in the K2K \cite{Catanesi:2005rc} and MiniBoone \cite{Catanesi:2007ab} targets,
 the NA61 experiment  \cite{Berns:2018tap} at CERN (see Fig. \ref{fig:na61}) has been extended beyond the CERN Long Shutdown 2 \cite{na61} to perform measurement campaigns with thin and replica targets of T2K, \nova, Hyper-K and DUNE.  

\subsection{Cross section experiments}

\begin{figure}
\centering
\includegraphics[width=0.6\textwidth]{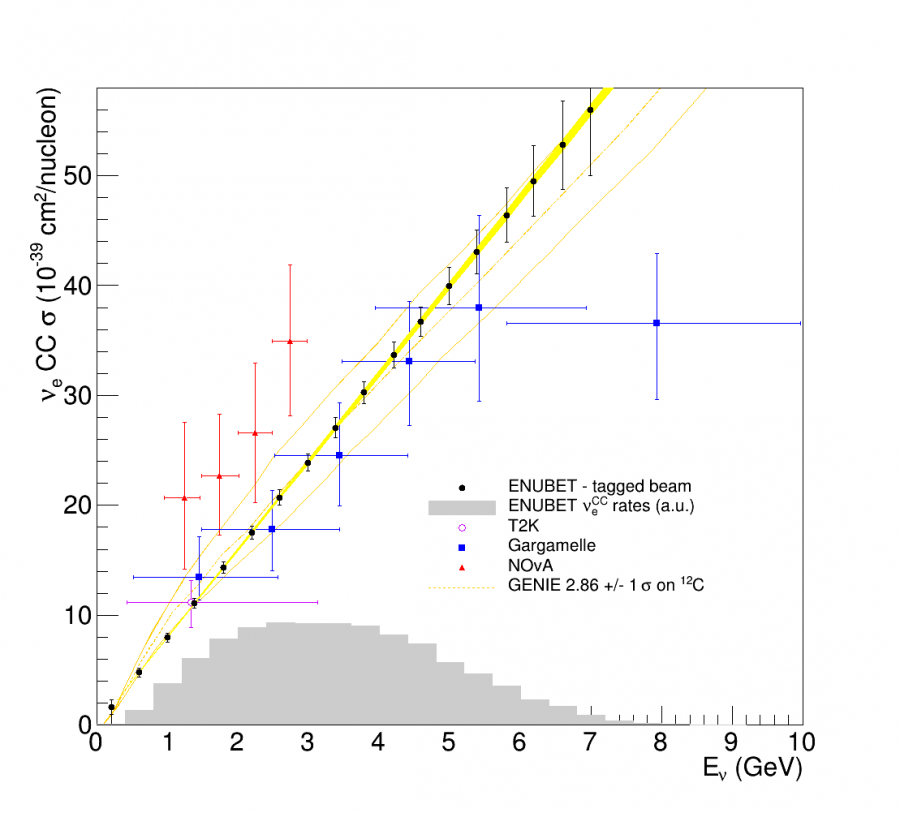}
\caption{Impact of ENUBET on the knowledge of the $\nu_e$ cross section. The ENUBET measurement and the statistical errors are shown with black dots. The orange bands represent the current systematic uncertainty from GENIE. The yellow band is the ENUBET systematic uncertainty on the flux. Past measurements from Gargamelle, T2K and \nova are also shown. The gray area shows the $\nu_e$ CC energy spectrum of ENUBET.}
\label{fig:enubet_xsect}
\end{figure}

In the last decade, our knowledge of total and differential cross sections at the GeV scale has grown significantly through dedicated experiments (SciBoone, Minerva, ArgoNeut), experiments operated in short baseline beams to search for sterile neutrinos (MiniBooNE, MicroBooNE), and the near detectors of short baseline experiments \cite{Alvarez-Ruso:2017oui,katori2018}. The results of these experiments showed significant disagreements with theory predictions and with the neutrino interaction generators used by long-baseline experiments. They hence played a prominent role in the systematic reduction programme of T2K and \nova. In the DUNE/Hyper-K era, their role will be even more important \cite{Huber:2007em,Coloma:2012ji}. In order to reduce the systematic budget at the per-cent level we cannot rely only on cancellation effects in the 
ratio
\begin{equation}
\frac{ \int dE \ \Phi_{\nu_e}(E) \sigma_{\nu_e}(E) \epsilon_{\nu_e}(E)  }{\int dE \ \Phi_{\nu_\mu}(E) \sigma_{\nu_\mu}(E) \epsilon_{\nu_\mu}(E) }  
\end{equation}
but we have to master each term of the integrals and, in particular $\sigma_{\nue}$ and $\sigma_{\numu}$ (currently known at the 25\% and 10\% level, respectively),  and the corresponding differential cross sections. Even if $\sigma_{\nue}$ is still statistics dominated and can be improved by employing the near detectors of DUNE and T2K-II, the dominant systematic contribution to the muon and electron neutrino cross sections will soon become the uncertainty on the flux of the short baseline beam used for the measurement of $\sigma_{e/\mu}$. This scenario advocates for a new generation of cross section experiments with a highly controlled beam. The most promising candidates are the "monitored neutrino beams" \cite{Longhin:2014yta} where the flux is measured in a direct manner monitoring the rate of leptons in the decay tunnel. In particular the CERN NP06/ENUBET \cite{ENUBET} experiment (ERC Consolidator Grant, PI A. Longhin) is aimed at designing a beamline where the positrons produced at large angles by $K^+ \rightarrow e^+ \pi^0 \nue$ decays are monitored at single particle level to measure the flux of \nue. Monitored neutrino beams may achieve a precision of <1\% in the flux determination and deliver $10^4$ ($10^6$) \nue (\numu) CC events per year in a 500 ton detector as ProtoDUNE-SP (CERN) or ICARUS (Fermilab). In addition, the ENUBET beam is a narrow band beam where the position of the neutrino interaction vertex in the neutrino detector is strongly correlated to the energy of the neutrino. It can therefore provide a 8\% (22\%) precision measurement of the energy of the incoming neutrino at 3 (1) GeV without relying on the kinematics of the final state particle. Energy reconstruction in total and differential cross sections is thus not biased by nuclear effects. The ENUBET beamline can be exploited not only by LArTPCs but also by dedicated high-granularity detectors specifically designed for cross section measurements at low and high Z. The impact of the ENUBET beamline on the measurement of the $\nu_e$ cross section is depicted in Fig.~\ref{fig:enubet_xsect}.

A much larger statistics of $\nue$ can be achieved by changing the core technology of accelerator neutrino beams. R\&D  on muon colliders and Neutrino Factories are ongoing since more than 30 years. In a Neutrino Factory \cite{Geer:1997iz,Choubey:2011zzq}, it is possible to produce \nue (\nubare) at the sama pace as \numu (\nubarmu) from the three body decay of muons stored in the accumulator ring. In its simplest design (NUSTORM \cite{Adey:2013pio,Adey:2015iha}), a low energy Neutrino Factory can deliver up to $10^6$ \nue CC events per year to the same detectors employed for ENUBET. As a consequence, NUSTORM would represent both a powerful source for cross section measurements and a major step forward in the design of facilities for the post-DUNE era (see Sec.\ref{sec:future}).

\section{A glimpse to the long-term future}
\label{sec:future}

It is hard to envisage how the field of accelerator neutrino beams will be shaped by the results of DUNE and Hyper-Kamiokande. A confirmation of the persistent sterile neutrino hints \cite{Giunti:2019aiy} would change in a substantial manner the three family oscillation paradigm considered in this paper and would urge a re-definition of the entire experimental strategy. Still, even in the framework of the minimally extended Standard Model, the Superbeams will not say the final word on lepton mixing. The PMNS precision in 2030 will still be very far from the precision achieved in the CKM. In flavor physics, the main tool to explore physics beyond the Standard Model are unitarity tests and inconsistencies between the mixing parameters measured by different channels: both tools require precision. In addition, the striking difference between the PMNS and CKM indicates that the entire Yukawa sector of the SM is not understood and that the origin of such difference may reside in physics at energy scales not accessible at accelerators. A per-cent level measurement of the PMNS, combined with the study of neutrino non-standard interactions (NSI) and the $\nue \rightarrow \nutau$ transitions (unitarity tests) are the main route toward these scales and require employing neutrino oscillations at an unprecedented level of precision.
Any upgrade of DUNE and Hyper-Kamiokande will be of little impact because at full statistics they will be limited by systematic errors and those errors cannot be further significantly reduced, being related to the intrinsic limitations of conventional neutrino beams.
New concepts in neutrino beams, as Neutrino Factories and/or Beta Beams will then become necessary for any further development after the third generation of long baseline neutrino experiments.
A bridge between the DUNE/Hyper-K era and the Neutrino Factory era will likely occur by upgrading existing facilities. In the European context a special role will be played by the European Spallation Source, the most powerful proton source devised for material science studies. The infrastructure and technology of ESS can be upgraded to serve the neutrino science community with a low energy Superbeam operating at the second oscillation maximum ($\sim$350 MeV) and, thus, with the largest CP sensitivity. The ESS also represents an interesting first stage facility for the implementation of the accelerator complex serving a Neutrino Factory and a high energy Muon Collider.       

\section{Conclusions}

In the last ten years accelerator neutrino beams consolidated their role in the study of neutrino oscillations. In this paper, we reviewed the most important experimental results provided by neutrino beam experiments and interpreted these results in the framework of the minimally extended Standard Model. Neutrino beams have established in a conclusive manner the  appearance of tau neutrinos (\numunutau $\ $oscillations) at CNGS and of electron neutrinos (\numunue $\ $oscillations) at J-PARC and NUMI.
T2K and \nova lead the measurements of the oscillation parameters at the atmospheric scale (\dmtt and \thetatt). In particular, the T2K+\nova combined measurement provide hints both of  normal hierarchy and of CP violation. These hints will be checked conclusively in the next generation of Superbeam, namely by DUNE and Hyper-Kamiokande. Our field is now in a stage where a superior control of systematics and a precision knowledge of neutrino interactions in matter is of paramount importance to ground the physics reach of the Superbeams. These needs foster a vigorous systematic reduction programme based on movable near detectors, hadroproduction experiments and a new generation of facilities for the study of cross sections at the GeV scale. 

\section*{Acknowledgements}
This work is supported by the EU H2020 programme under grant agreement n.~674896 and n.~777419,  by the COST action CA15139 and by RISE-GA822070-JENNIFER2 2020


\reftitle{References}





\end{document}